\documentclass{elsarticle}
\usepackage{ifpdf}
\RequirePackage[center]{subfigure} 
\usepackage{graphicx,amssymb,lineno}
\usepackage[english]{babel}
\usepackage[latin1]{inputenc}

\graphicspath{{Figures/}}
\ifpdf
\usepackage[%
  pdftitle={QUATERNIONS
    elsart},%
  pdfauthor={Arnaud Lazarus},%
  pdfsubject={Quaternions Asymptotic Numerical Method for the continuation of equilibriums of elastic rods},%
  pdfkeywords={Asymptotic numerical continuation method, quaternions method, equilibriums and stability under constrains, elastic rods},%
  pdfstartview=FitH,%
  bookmarks=true,%
  bookmarksopen=true,%
  breaklinks=true,%
  colorlinks=true,%
  linkcolor=blue,anchorcolor=blue,%
  citecolor=blue,filecolor=blue,%
  menucolor=blue,pagecolor=blue,%
  urlcolor=blue]{hyperref}
\else
\usepackage[%
  breaklinks=true,%
  colorlinks=true,%
  linkcolor=blue,anchorcolor=blue,%
  citecolor=blue,filecolor=blue,%
  menucolor=blue,pagecolor=blue,%
  urlcolor=blue]{hyperref}
\fi

\makeatletter
\def\elsartstyle{%
    \def\normalsize{\@setfontsize\normalsize\@xiipt{14.5}}
    \def\small{\@setfontsize\small\@xipt{13.6}}
    \let\footnotesize=\small
    \def\large{\@setfontsize\large\@xivpt{18}}
    \def\Large{\@setfontsize\Large\@xviipt{22}}
    \skip\@mpfootins = 18\p@ \@plus 2\p@
    \normalsize
}
\@ifundefined{square}{}{}
\makeatother

\DeclareMathAlphabet{\mathdj}{U}{msb}{m}{n}
\newcommand{\vect}[1]{\bm{#1}}
\newcommand{\udots}{\mathinner{\mskip1mu\raise1pt\vbox{\kern7pt\hbox{.}}
  \mskip2mu\raise4pt\hbox{.}\mskip2mu\raise7pt\hbox{.}\mskip1mu}}

\newcommand{\mat}[1]{\ensuremath{\bm{#1}}}

\newcommand{\tp}[1]{{#1}^T}

\usepackage{amsmath,amsfonts,bm,empheq}
\usepackage[normalem]{ulem}
\usepackage{booktabs}
\usepackage{tabularx}
\usepackage{multirow}
\usepackage{rotating}
\usepackage{xspace}
\usepackage{esint}
\usepackage{url}
\usepackage{array,hhline}
\usepackage{color}
\usepackage{amsthm}

\definecolor{darkgreen}{rgb}{0,0.6,0}
\newcommand{\colF}[1]{{\color{darkgreen} #1}}

\begin{document}
\begin{frontmatter}

\title{A quaternion-based continuation method to follow the equilibria and stability of slender elastic rods}

\author[us]{A. Lazarus}
\ead{alazarus@mit.edu}
\author[him]{J. T. Miller}
\ead{flymile@mit.edu}
\author[us,him]{P. M. Reis}
\ead{preis@mit.edu}

\address[us]{Department of Mechanical Engineering, 
Massachusetts Institute of Technology,
Cambridge 02139, USA}
\address[him]{Department of Civil and Environmental Engineering, 
Massachusetts Institute of Technology,
Cambridge 02139, USA}

\begin{abstract}
We present a theoretical and numerical framework to compute bifurcations of equilibria and stability of slender elastic rods. The 3D kinematics of the rod is treated in a geometrically exact way by 
parameterizing the position of the centerline and making use of quaternions to represent the orientation of the material frame. The equilibrium equations and the stability of their solutions are derived from the mechanical energy which takes into account the contributions due to internal moments (bending and twist), external forces and torques.  Our use of quaternions allows for the equilibrium equations to be written in a simple quadratic form and solved efficiently with an asymptotic numerical continuation method. This finite element perturbation method gives interactive access to semi-analytical equilibrium branches, in contrast with the individual solution points obtained from classical minimization or predictor-corrector techniques. By way of example, we apply our numerics to address the specific problem of a naturally curved rod under extreme twisting and perform a detailed comparison against our own precision model experiments of this system. Excellent quantitative agreement is found between experiments and simulations for the underlying 3D buckling instabilities and the characterization of the resulting complex configurations.
\end{abstract}

\end{frontmatter}

\newpage
\section{Introduction}
\label{intro}

Filaments, rods and cables are encountered over a wide range of length-scales, both in nature and technology, providing outstanding kinematic freedom for practical applications. Given their slender geometry, they can undergo large deformations and exhibit  complex mechanical behavior including  buckling, snap-through and localization.
A predictive understanding of the mechanics of thin rods has therefore long motivated a large body of theoretical and computational work, from Euler's elastica in $1744$ \cite{levien2008elastica} and Kirchhoff's kinetic analogy in $1859$ \cite{dill1992kirchhoff} to the burgeoning of numerical approaches such as finite element-based methods in the late $20^{\text{th}}$ century \cite{zienkiewicz2005finite}, and the more recent algorithms based on discrete differential geometry \cite{bergou2008discrete2}. Today, these advances in modeling of the mechanics of  slender elastic rods are helping to tackle many cutting-edge research problems. To name just a few, these range from the supercoiling of DNA  \cite{Marko2012Competition}, self-assembly of rod-coil block copolymers \cite{wang2012diffusion}, design of nano-electromechanical resonators \cite{lazarus2010statics,lazarus2010simple}, development of stretchable electronics \cite{sun2006controlled}, computed animation of hairs \cite{bertails2006super} and coiled tubing operations in the oil-gas industries \cite{wicks2008horizontal}.

An ongoing challenge in addressing these various problems involves the capability to numerically capture their intrinsic geometric nonlinearities in a predictive and efficient way. These nonlinear kinematic effects arise from the large displacements and rotations of the slender structure, even if its material properties remain linear throughout the process \cite{audoly2000elasticity}. As a slender elastic rod is progressively deformed, the nonlinearities of the underlying equilibrium equations become increasingly stronger leading to higher densities in the landscape of possible solutions for a particular set of control parameters. When multiple stable states coexist, classic step-by-step algorithms such as Newton-Raphson methods \cite{crisfield1991nonlinear} or standard minimization techniques \cite{luenberger1973introduction} are often inappropriate since, depending on the initial guess, they may not converge towards the desired solution, or any solution.  Addressing these computational difficulties calls for alternative numerical techniques, such as well-known continuation methods \cite{riks1979incremental,koiter1970stability}. Continuation techniques are based on coupling nonlinear algorithms (e.g. predictor-corrector \cite{riks1979incremental} or perturbation methods \cite{koiter1970stability}) with an arc-length description to numerically follow the fixed points of the equilibrium equations as a function of a control parameter, that is often a mechanical or geometrical variable of the problem. With the goal of determining the complete bifurcation diagram of the system, these methods enable the computation of all of the equilibrium solution branches, as well as their local stability.   

Two main approaches can be distinguished for continuing the numerical solutions of geometrically nonlinear problems. The first includes predictor-corrector methods whose principle is to follow the nonlinear solution branch in a stepwise manner, via a succession of linearizations and iterations to achieve equilibrium \cite{crisfield1991nonlinear}. These methods are now widely used, particularly for the numerical investigation of solutions of conservative dynamical systems, with the free path-following mathematical software AUTO being an archetypal example \cite{doedel1981auto}. Quasi-static deformations of slender elastic rods have been intensively studied using this software \cite{thompson1996helix,healey2005straightforward}, mostly due to the analogy between the rod's equilibrium equations with the spinning top's dynamic equations \cite{davies3Dspatial1993}. Although popular and widely used, the main difficulty with these algorithms involves the determination of an appropriate arc-length step size, which is fixed \emph{a priori} by the user, but may be intricately dependent on the system's nonlinearities along the bifurcation diagram. A smaller step size will favor the computation of the highly nonlinear part of the equilibrium branch, such as bifurcation points, but may also impractically increase the overall computational time. On the other hand, a larger step size may significantly compromise the accuracy and resolution of the results.

The second class of continuation algorithms, which have received less attention, is a perturbation technique called the \emph{Asymptotic Numerical Method} (ANM), which was first introduced in the early 1990's \cite{damil1990new,cochelin1994path}. The underlying principle is to follow a nonlinear solution branch by applying the ANM in a stepwise manner and represent the solution by a succession of local polynomial approximations. This numerical method is a combination of asymptotic expansions and finite element calculations which allows for the determination of an extended portion of a nonlinear branch at each step, by inverting a unique stiffness matrix. This continuation technique is significantly more efficient than classical predictor-corrector schemes. Moreover, by taking advantage of the analytical representation of the branch within each step, it is highly robust and can be made fully automatic. Unlike incremental-iterative techniques, the arc-length step size in ANM is adaptative since it is determined \emph{a posteriori} by the algorithm. As a result, bifurcation diagrams can be naturally computed in an optimal number of iterations. The method has been successively applied to nonlinear elastic structures such as beams, plates and shells but the geometrical formulations were limited to the early post-buckling regime and to date, no stability analyses were performed with ANM \cite{cochelin1994asymptotic,zahrouni1999computing,vannucci1998asymptotic}.

In this paper, we develop a novel implementation of the semi-analytical ANM algorithm to follow the equilibrium branches and local stability of slender elastic rods with a geometrically-exact 3D kinematics. In Section \ref{sec2}, we first describe the 3D kinematics where the rod is represented by the position of its centerline and a set of unit quaternions to represent the orientation of the material frame. In Section \ref{sec3}, we then derive the closed form of the rod's nonlinear equilibrium equations, by minimizing its geometrically-constrained mechanical energy including internal bending and twisting energy, as well as the work of external forces and moments. Expressing the flexural and torsional internal moments in a quaternion basis yields differential equilibrium equations that are simply quadratic in terms of the unknowns. In Section \ref{sec4}, we proceed by presenting the numerical method developed to compute the equilibrium solutions. Using a finite element approach, the discretized system of equilibrium equations can be solved with the ANM algorithm, which is particularly efficient for computing our algebraic quadratic form. The local stability of the computed equilibrium branches is assessed by a second order condition on the constrained energy. Finally, we describe how to implement this numerical method in the open source software MANlab; a user-friendly, interactive and Matlab-based path-following and bifurcation analysis program \cite{arquier2007methode,manlab}. In Section \ref{sec5}, we develop our own precision model experiment for the fundamental problem of the writhing of a clamped elastic rod \cite{thompson1996helix,goriely1998nonlinear4, van2000helical,goyal2008non} and challenge our numerical model against experimental results. The simulations are robust, computationally time-efficient and exhibit excellent quantitative agreements with our experiments, demonstrating the predictive power of our framework. 

\section{Kinematics}\label{sec2}

In this section, we present the formulation for the geometry and 3D kinematics of the slender elastic rod that we will use in our study. Assuming no shear strains and inextensibility, the mechanical deformations are represented by the rate of change of the orientation along the rod,  characterized by a set of geometrically constrained unit quaternions.   

\begin{figure}[b!]
\includegraphics[width=\columnwidth]{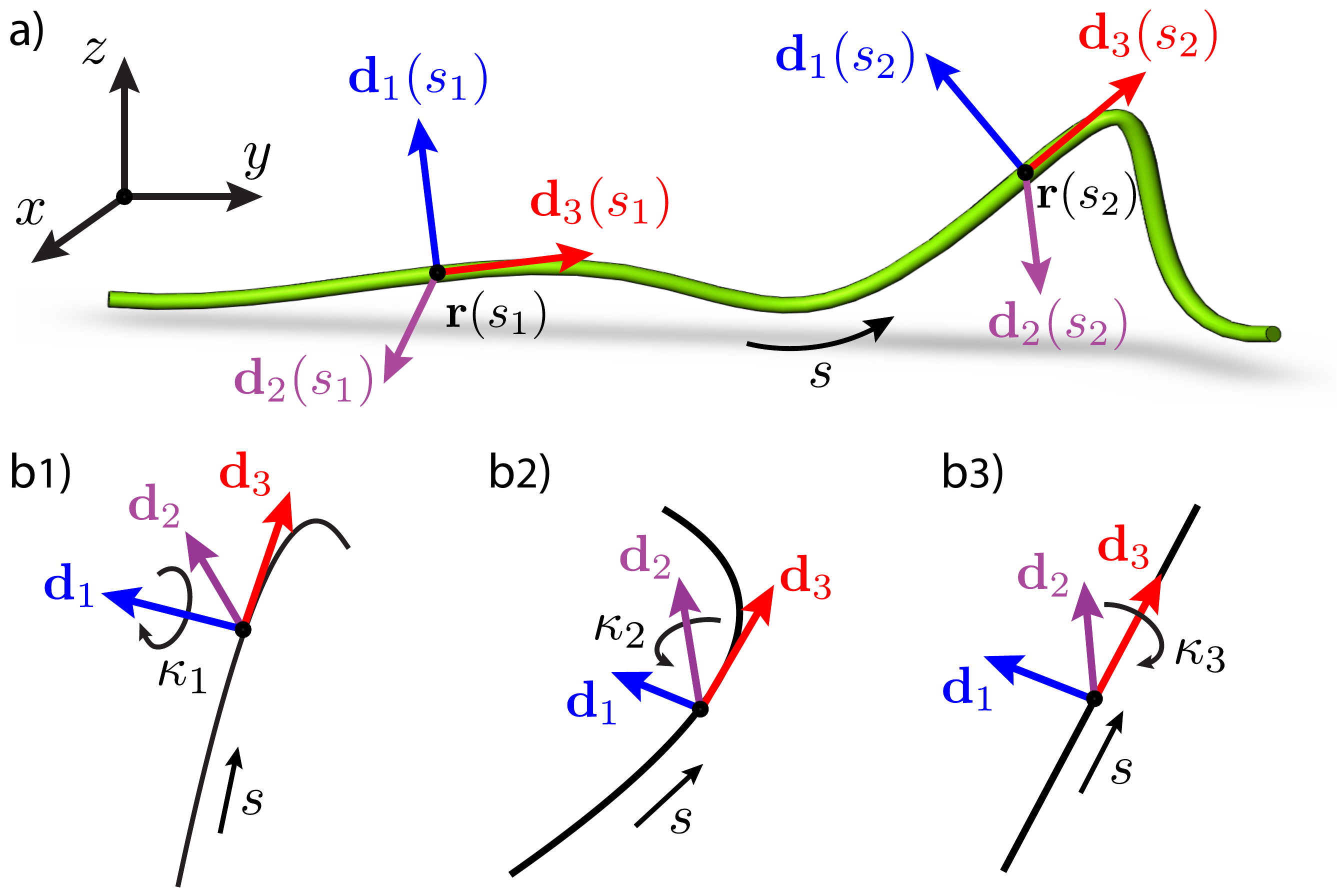}
\caption{\label{fig:Cosserat_rod} Kinematics of the Cosserat rod in the global cartesian frame $\left(x,y,z\right)$. (a) The configuration of the rod is defined by its centerline $\vect{r}(s)$. The orientation of each mass point of the rod is represented by an orthonormal basis $\left(\vect{d}_1(s) \, \vect{d}_2(s) \, \vect{d}_3(s)\right)$, called the directors, where $\vect{d}_3(s)$ is constrained to be tangent to $\vect{r}(s)$. (b) The three local modes of deformation of the elastic rod, associated with the change of (b1) material curvature $\kappa_1$ related to the direction $\vect{d}_1$ of the cross-section, (b2) material curvature $\kappa_2$ related to $\vect{d}_2$, and (b3) twist. }
\end{figure}

\subsection{Cosserat theory of elastic rods}

An elastic rod is a slender elastic body which has a length along one spatial direction that is much larger than its dimensions in the two other perpendicular directions, that define the cross section [Fig.~\ref{fig:Cosserat_rod}(a)]. We denote the typical size of the cross-section by $h$ and the  other length scale by $L$. At large scales, the rod can be regarded as an adapted material curve: its centerline. If $s$ denotes the curvilinear coordinate along the centerline of the undeformed rod, we can represent this line by a position vector function (with respect to some fixed origin) of the material point originally at $s$ in the reference configuration,
\begin{equation}  
\vect{r}(s) =  \tp{\left[r_x(s) \, r_y(s) \, r_z(s)\right]} = \tp{\left[x(s) \, y(s) \, z(s)\right]}.
\label{eqpositioneuclidean}
\end{equation}  
We consider unstretchable rods whose centerline remains inextensible upon deformation. As explained in detail in \cite{audoly2000elasticity}, this assumption is physically justified for a wide range of loading conditions, provided that the aspect ratio of the rod, $h/L$, is small. Under this assumption, the variable $s$ is also the curvilinear coordinate along the centerline in the actual configuration. The configuration of the rod is not only characterized by the path of its centerline but also by how much it twists around this line. We consider this twist by introducing the material frame $\left(\vect{d}_1(s) \, \vect{d}_2(s) \, \vect{d}_3(s)\right)$ in the deformed configuration. At each particular location $s$, we associate an orthonormal basis $\vect{d}_k$, $(k = 1,2,3)$ attached to the centerline. The centerline, together with this set of material frames, form what is called a Cosserat curve \cite{spillmann2009cosserat}. We choose the orientation of these material frames in a way such that the directors $\vect{d}_1$ and $\vect{d}_2$ lie in the plane of the cross-section, while the third director $\vect{d}_3$ is always parallel to the tangent of the curve [see Fig.~\ref{fig:Cosserat_rod}(a)]. Considering the case of small strains, the triad $\left(\vect{d}_1(s) \, \vect{d}_2(s) \, \vect{d}_3(s)\right)$ remains approximately orthonormal upon deformation. This is known as the Euler-Bernoulli kinematical hypothesis (assumption of no shear deformations).  

Before we are able to establish the constitutive relation, we have to quantify the rate of change of position and orientation along the rod's centerline. The rate of change in the position of the centerline is a strain vector $\vect{v} (s)= \tp{\left[v_1(s) \, v_2(s) \, v_3(s)\right]}$ that vanishes since shearing in both transverse directions and stretching are neglected. Therefore, the strains arise from the orientational rate of change of the cross-sections alone, which we now express using the framework of differential geometry of curves in 3D space \cite{audoly2000elasticity}. The previous condition of orthonormality (Euler-Bernoulli assumption) yields the relations,
\begin{equation}  
\vect{d}'_i(s).\vect{d}_i(s) = 0 \quad \textnormal{and} \quad \vect{d}'_i(s).\vect{d}_j(s) = -\vect{d}'_j(s).\vect{d}_i(s),
\label{eqorthonormality}
\end{equation}
for all indices $i$ and $j$ varying from $1$ to $3$ (there is no implicit sum over $i$ in the first equation) and where $(\,)'$ denotes differentiation with respect to $s$. The most general set of first-order linear equations conserving the orthonormal character of the material frame $\left(\vect{d}_1(s) \, \vect{d}_2(s) \, \vect{d}_3(s)\right)$ represented in Eq.~(\ref{eqorthonormality}) can be expressed as,
\begin{subequations}
\label{eqdirectorsderivative}
\begin{equation}
\vect{d}'_1(s)= \kappa_3(s)\vect{d}_2(s) - \kappa_2(s)\vect{d}_3(s), \end{equation}
\begin{equation}
\vect{d}'_2(s)= -\kappa_3(s)\vect{d}_1(s) + \kappa_1(s)\vect{d}_3(s),
\end{equation} 
\begin{equation}
\vect{d}'_3(s)= \kappa_2(s)\vect{d}_1(s) - \kappa_1(s)\vect{d}_2(s).
\end{equation} 
\end{subequations}
where $\kappa_1(s)$, $\kappa_2(s)$ and $\kappa_3(s)$ are scalar functions interpreted below. These equations describe the rigid-body rotation of the frame. Using the notation $\vect{u} \times \vect{v}$ for the cross product of two vectors, Eqs.~(\ref {eqdirectorsderivative}) can be rewritten as, 
\begin{equation}  
\vect{d}'_1(s) = \vect{\Omega}(s) \times \vect{d}_1(s), \quad \vect{d}'_2(s) = \vect{\Omega}(s) \times \vect{d}_2(s), \quad \vect{d}'_3(s) = \vect{\Omega}(s) \times \vect{d}_3(s),
\label{eqdirectorsderivativebis}
\end{equation}
where we have introduced the Darboux vector $\vect{\Omega}(s)$,
\begin{equation}  
\vect{\Omega}(s) = \kappa_1(s) \vect{d}_1(s) + \kappa_2(s) \vect{d}_2(s) + \kappa_3(s) \vect{d}_3(s).
\label{eqdarbouxvector}
\end{equation}
The physical interpretation of Eqs.~(\ref{eqdirectorsderivativebis}) is that the material frame rotates with a \emph{rotation velocity}, $\vect{\Omega}(s)$, when following the centerline at unit speed. The quantities $\kappa_1$ and $\kappa_2$ in Eq.~(\ref{eqdarbouxvector}), called the material curvatures, illustrated in Fig.~\ref{fig:Cosserat_rod}(b1)-(b2), represent the extent of rotation of the material frame, with respect to the directions $\vect{d}_1$ and $\vect{d}_2$ of the cross-section. The quantity $\kappa_3$ quantifies the rotation of the material frame with respect to the tangent $\vect{d}_3$, and is called the material twist of the rod [see Fig.~\ref{fig:Cosserat_rod}(b3)]. In order to write the material curvature and twist in an explicit form, the Darboux vector has to be rotated into the local frame. Using the condensed notation, $\vect{\kappa}(s) = \tp{\left[\kappa_1(s) \, \kappa_2(s) \, \kappa_3(s)\right]}$, and the rotation matrix of the Euclidean 3D space $\mat{R}(s) \in \mathcal{R}^{3\times 3}$, this rotated Darboux vector is,
\begin{equation}  
\vect{\kappa}(s) = \tp{\mat{R}}(s)\vect{\Omega}(s),
\label{eqbendingtwistrotation}
\end{equation}
or, in terms of the directors $\vect{d}_k(s)$,
\begin{equation}  
\kappa_k(s) = \vect{d}_k(s).\vect{\Omega}(s),
\label{eqbendingtwistdirectors}
\end{equation}
since the directors $\vect{d}_k(s)$ constitute the columns of the rotation matrix $\mat{R}(s) = [\vect{d}_1(s) \, \vect{d}_2(s) \, \vect{d}_3(s)]$. 

The 3D kinematics formulation of our inextensible and unshearable elastic rod is not yet complete because we won't be able to derive the equilibrium equations directly from the material curvatures. In fact, a difficulty arises when trying to compute the infinitesimal work of the external forces using the variables $\kappa_1$, $\kappa_2$ and $\kappa_3$. A perturbation of these quantities yields a non-local perturbation to the centerline and attached material frame so that the work of the external forces cannot be written in a straightforward manner \cite{audoly2000elasticity}. Instead, the classic approach is to choose as degrees of freedom the orientation of the material frame characterized, in this paper, by a set of quaternions. We shall now explain how to represent the rotation matrix $\mat{R}(s)$ or the directors, $\vect{d}_k(s)$, and the strain rate vector, $\vect{\kappa}(s)$, in the framework of quaternions.

\subsection{Quaternion representation}

Quaternions are a number system that extends the complex number representation of geometry in a plane to the three-dimensional space \cite{altmann1986rotations}. They were first described by Hamilton in 1843 \cite{hamilton1847quaternions,hamilton1853lectures} and were extensively used in many physics and geometry problems before loosing prominence in the late $19^{\text{th}}$ century following the development of numerical analysis. Quaternions were then revived in the late $20^{\text{th}}$ century, primarily due to their power and simplicity in describing spatial rotations, and have since been revived in a wide range of fields: applied mathematics \cite{kuipers1999quaternions}, computer graphics \cite{dam1998quaternions,hanson2005visualizing}, optics \cite{horn1987closed,tweed1990computing}, robotics \cite{chou1991finding} and orbital mechanics \cite{arribas2006quaternions,waldvogel2008quaternions}. It is beyond the scope of this article to discuss a detailed evaluation of the advantages and disadvantages of using quaternions over other rotation parametrizations. However, we highlight that quaternions are a non-singular representation of rotation, unlike Euler angles for instance, even if they are less intuitive than direct angles. Moreover, we favor quaternions over trigonometric approaches because of their remarkably compact quadratic polynomial form. We will show that one striking outcome of using quaternions is that the equilibrium equations we shall derive are, at most, cubic in terms of the degrees of freedom. This property is at the heart of the numerical continuation method presented in Section \ref{sec4}.   

The fundamental relation of the algebra of quaternions, denoted by $\mathcal{H}$, is,
\begin{equation}  
i^2 = j^2 = k^2 = ijk = -1,
\label{eqformulaquaternions}
\end{equation} 
where $i$, $j$, and $k$ are the basis elements of $\mathcal{H}$. A quaternion number in $\mathcal{H}$ is written in the form $q_1i + q_2j + q_3k + q_4$ where the imaginary part $q_1i + q_2j + q_3k$ is an element of the vector space $\mathcal{R}^3$ and the real part $q_4$ is a scalar. Using the basis $i$, $j$, $k$, $1$ of $\mathcal{H}$ makes it possible to write a quaternion as a set of quadruples, usually expressed as a vector in $\mathcal{R}^4$,
\begin{equation}  
\vect{q} = \tp{\left[q_1 \, q_2 \, q_3 \,q_4\right]}.
\label{eqquaternionvector}
\end{equation}

\begin{figure}[t!]
\includegraphics[width=\columnwidth]{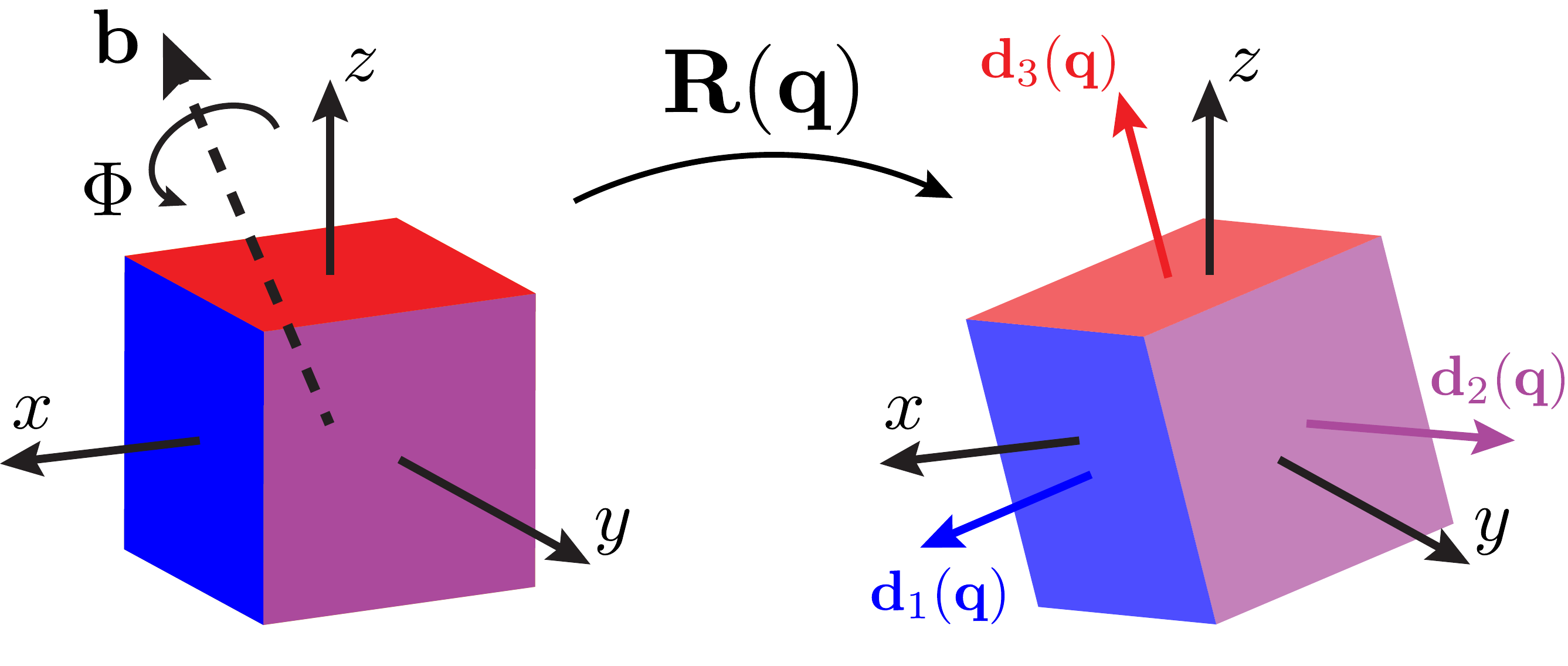}
\caption{\label{fig:quaternions} Rotation of a rigid body using Euler's rotation theorem and a set of unit quaternions. Knowing the rotation angle $\phi$ around the unit vector $\vect{b}$, we can associate a rotation matrix $\mat{R}(\vect{q})$ and three directors $\vect{d}_1(\vect{q})$, $\vect{d}_2(\vect{q})$ and $\vect{d}_3(\vect{q})$ expressed exclusively in terms of quaternions according to Eqs.~(\ref{eqeulerparameter})-(\ref{eqrotationmatrixdirectorsquaternion}).}
\end{figure}

Quaternions of norm one, or \emph{unit quaternions}, are a particularly  convenient mathematical notation for representing orientations of objects in three dimensions. Using Euler's rotation theorem which states that a general re-orientation of a rigid-body can be accomplished by a single rotation about some fixed axis, one can represent a rotation by a set of quaternions, known as Euler parameters,
\begin{equation}  
\vect{q} = \tp{\left[b_x\!\sin(\Phi/2) \; b_y\!\sin(\Phi/2) \; b_z\!\sin(\Phi/2) \; \!\cos(\Phi/2)\right]},
\label{eqeulerparameter}
\end{equation}       
where $\Phi$ is the Euler principal angle and $\vect{b} =  \tp{\left[b_x \, b_y \, bz\right]}$ is the unit length principal vector such that $b_x^2 + b_y^2 + b_z^2 =1$ [see Fig.~\ref{fig:quaternions}]. Given that four Euler parameters are needed to define a three-dimensional rotation, a natural constraint equation prescribing that $\vect{q}$ is indeed a unit quaternion follows from Eq.~(\ref{eqeulerparameter}),
\begin{equation}  
q_1^2 + q_2^2 + q_3^2 + q_4^2 = 1.
\label{equnitquaternion}
\end{equation} 
The orthogonal matrix representation corresponding to a rotation by the quaternion $q = q_1i + q_2j + q_3k + q_4$ with $\lVert q\rVert = 1$ is, 
\begin{equation}  
\mat{R}(\vect{q}) = \left[ \begin{array}{ccc}
q_1^2 - q_2^2 - q_3^2 + q_4^2 & 2(q_1q_2 - q_3q_4) & 2(q_1q_3 + q_2q_4) \\
2(q_1q_2 + q_3q_4) & -q_1^2 + q_2^2 - q_3^2 + q_4^2 & 2(q_2q_3 - q_1q_4) \\
2(q_1q_3 - q_2q_4) & 2(q_2q_3 + q_1q_4) & -q_1^2 - q_2^2 + q_3^2 + q_4^2 \end{array} \right].
\label{eqrotationmatrixquaternion}
\end{equation}

Returning to the context of a thin elastic rod discussed above, its material frame $\left(\vect{d}_1(s) \, \vect{d}_2(s) \, \vect{d}_3(s)\right)$ remains orthonormal upon deformations and its rigid-body re-orientation can be expressed by the rotation matrix given in Eq.~(\ref{eqrotationmatrixquaternion}) such that, 
\begin{equation}  
\mat{R}(\vect{q}(s)) = \left[\vect{d}_1(\vect{q}(s)) \; \vect{d}_2(\vect{q}(s)) \; \vect{d}_3(\vect{q}(s))\right].
\label{eqrotationmatrixdirectorsquaternion}
\end{equation}
The local frame is now parametrized in terms of the curvilinear unit quaternion coordinates vector $\vect{q}(s) = \tp{\left[q_1(s) \, q_2(s) \, q_3(s) \,q_4(s)\right]}$ along the slender rod. 

We proceed by relating the strain rate vector $\vect{\kappa}(s)$ of Eq.~(\ref{eqbendingtwistrotation}) to the Euler parameters $\vect{q}(s)$. Multiplying each of the three geometric relations given in Eqs.~(\ref{eqdirectorsderivative}) by the relevant director $\vect{d}_k$ yields expressions for the material curvatures and twist in terms of the directors alone,
\begin{equation}  
\kappa_1(s) = -\vect{d}_2(s).\vect{d}'_3(s), \quad \kappa_2(s) = -\vect{d}_3(s).\vect{d}'_1(s), \quad \kappa_3(s) = -\vect{d}_1(s).\vect{d}'_2(s).
\label{eqstrainratefctdirectors}
\end{equation}
To compute $\vect{d}'_k(s)$ in terms of quaternions, we note that $\vect{d}'_k(s)$ is a function of $\vect{q}(s)$, which is itself a function of the curvilinear coordinate $s$. Upon employing the chain rule of partial differentiation, we obtain
\begin{equation}  
\vect{d}'_k(s) = \vect{d}'_k(\vect{q}(s)) = \frac{\partial \vect{d}_k(\vect{q}(s))}{\partial s} = \frac{\partial\vect{d}_k}{\partial\vect{q}} \frac{\partial\vect{q}}{\partial s} = \mat{J}_k(\vect{q}(s))\vect{q}'(s)
\label{eqderivativedirectors}
\end{equation}
for the three directions $k = 1,2,3$, where $\mat{J}_k(\vect{q}(s))$ is the Jacobi matrix $\mat{J}_k = \partial\vect{d}_k / \partial\vect{q}$. Replacing $\vect{d}_k(s)$ and $\vect{d}'_k(s)$ by their respective expressions (\ref{eqrotationmatrixquaternion})-(\ref{eqrotationmatrixdirectorsquaternion}) and  (\ref{eqderivativedirectors}) in Eq.~(\ref{eqstrainratefctdirectors}) allows us to express the material curvatures, $\kappa_1(s)$ and $\kappa_2(s)$, and the twist, $\kappa_3(s)$, solely in terms of the unit quaternions,
\begin{equation}  
\kappa_k(s) = 2\mat{B}_k\vect{q}(s)\vect{q}'(s) \quad \textnormal{for k=}1, 2, 3,
\label{eqstrainratesfinal}
\end{equation}
where the skew-symmetric matrices $\mat{B}_k$ read
\begin{gather}  
\mat{B}_1 = \left[ \begin{array}{cccc}
0 & 0 & 0 & 1 \\
0 & 0 & 1 & 0 \\
0 & -1 & 0 & 0 \\
-1 & 0 & 0 & 0 \end{array} \right], \quad \mat{B}_2 = \left[ \begin{array}{cccc}
0 & 0 & -1 & 0 \\
0 & 0 & 0 & 1 \\
1 & 0 & 0 & 0 \\
0 & -1 & 0 & 0 \end{array} \right], \quad
\mat{B}_3 = \left[ \begin{array}{cccc}
0 & 1 & 0 & 0 \\
-1 & 0 & 0 & 0 \\
0 & 0 & 0 & 1 \\
0 & 0 & -1 & 0 \end{array} \right].
\label{eqbk}
\end{gather}
With the new expression of $\kappa_k(s)$ in Eq.~(\ref{eqstrainratesfinal}), we have been able to write the total strains in terms of the locally perturbable variables $\vect{q}(s)$, which will be used in the derivation of the equations of equilibrium by variation of elastic energy presented below in Section \ref{sec3}. 

It is important to note, however, that the kinematic formulation is not yet complete since the four quaternions $q_1(s)$, $q_2(s)$, $q_3(s)$ and $q_4(s)$ are not geometrically independent. First, to represent a three-dimensional rotation with four coordinates, the unit quaternion assumption $\lVert\vect{q}(s)\rVert = 1$ given in Eq.~(\ref{equnitquaternion}), must be verified. Secondly, whereas thus far we have treated the centerline position $\vect{r}(s)$ and the orientations $\vect{q}(s)$ as separate entities, the positions and the orientations cannot be considered independently. Indeed, the material frames parametrized by $\vect{r}(s)$ and $\vect{q}(s)$ are coupled by the constraint that the third director $\vect{d}_3(\vect{q}(s))$ is always parallel to the tangent $\vect{r}'(s)$,
\begin{equation}  
\vect{r}'(s) = \vect{d}_3(\vect{q}(s)),
\label{eqadaptedframeconstraint}
\end{equation}
where $\vect{r}'(s)$ is the unit tangent vector to the Cosserat curve and $\|\vect{r}'(s)\| = 1$ along the centerline since we assumed inextensibility. The three constraints set by Eq.~(\ref{eqadaptedframeconstraint}) assure that the directors are adapted to the Cosserat curve [see Fig.~\ref{fig:Cosserat_rod}]. 

The three-dimensional kinematics of our inextensible and unshearable rod (including bending and twist) is represented by Eq.~(\ref{eqstrainratesfinal}), which links the strain rates to the local orientation of the material frame, together with the four geometrical constraints given in Eq.~(\ref{equnitquaternion}) and Eq.~(\ref{eqadaptedframeconstraint}). For the remainder of this article, the three positions $r_x(s), r_y(s), r_z(s)$ and four quaternion coordinates $q_1(s), q_2(s), q_3(s), q_4(s)$ constitute the seven degrees of freedom of our slender elastic rod \cite{spillmann2009cosserat,spillmann2010inextensible}. After taking into account the four constraint equations, only three of the DOFs are, in fact, geometrically independent. Their values are determined by the three-dimensional equilibrium equations, which we now address in the following section.

\section{Mechanical equilibrium}\label{sec3}

Having formulated the kinematics of our system, we proceed by analyzing the energetics of an arbitrary configuration of the slender elastic rod. We will then derive the equations for equilibrium obtained under the assumption that this energy is stationary under small deformations for the given boundary conditions and geometrical constraints introduced above. We highlight the fact that the equilibrium equations are highly nonlinear due to geometry, rather than the material response.

\subsection{Energy formulation}

For simplicity, and to avoid loss of generality, we shall adopt the framework of Hookean elasticity and consider linear isotropic constitutive laws. For practical purposes, this hypothesis is usually appropriate since, for slender elastic rods, the strains at the material level are typically small. Under this assumption, the total elastic energy of the slender elastic rod can be written as the uncoupled sum of bending and twisting contributions \cite{audoly2000elasticity}. Although the reference configuration of the rod is assumed to be stress-free, we can readily account for rods with intrinsic natural curvature and twist. Doing so, the elastic energy of a rod with length $L$ and a constant cross-section reads,
\begin{multline}  
\mathcal{E}_{e} = \frac{EI_1}{2}\int_0^L\left(\kappa_1(s) - \hat{\kappa}_1(s)\right)^2ds + \frac{EI_2}{2}\int_0^L\left(\kappa_2(s) - \hat{\kappa}_2(s)\right)^2ds \\
+ \frac{GJ}{2}\int_0^L\left(\kappa_3(s) - \hat{\kappa}_3(s)\right)^2ds,
\label{eqbendingenergy}
\end{multline}
where we used the previously defined rotational strain rate vector $\vect{\kappa}(s) = \tp{\left[\kappa_1(s) \, \kappa_2(s) \, \kappa_3(s)\right]}$, and where the quantities $\hat{\kappa}_1(s)$, $\hat{\kappa}_2(s)$ and $\hat{\kappa}_3(s)$  are the intrinsic natural curvature and twist of the rod along the directors $\vect{d}_1$,  $\vect{d}_2$ and $\vect{d}_3$, respectively. In this expression, $E$ is the Young's modulus of the material and $G = E/2(1+\nu)$ is the shear modulus of the material with Poisson's ratio $\nu$. The constants $I_1$ and $I_2$ are the moments of inertia along the principal directions of curvature in the plane of the cross-section $\vect{d}_1$ and $\vect{d}_2$ and $J$ is the moment of twist which, similarly to $I_1$ and $I_2$ for the bending energy, depends only of the geometry of the cross-section. Replacing the material curvatures $\kappa_1(s)$, $\kappa_2(s)$ and twist $\kappa_3(s)$ by their expression given in Eq.~(\ref{eqstrainratesfinal}) allows us to write the elastic energy $\mathcal{E}_{e}$ in a more compact form, in term of the rotational degrees of freedom $\vect{q}(s)$ alone,
\begin{equation}  
\mathcal{E}_{e}\left(\vect{q}(s)\right) = \sum_{k=1}^3\frac{E_kI_k}{2}\int_0^L\left(2\mat{B}_k\vect{q}(s)\vect{q}'(s) - \hat{\kappa}_k(s)\right)^2ds,
\label{eqbendingenergyquaternions}
\end{equation}
where $E_1 = E_2 = E$, $E_3 = G$ and $I_3 = J$.

\subsection{Variation of the energy}

We now follow a variational approach for the elastic energy in Eq.~(\ref{eqbendingenergyquaternions}), and consider an infinitesimal perturbation from an arbitrary configuration of the rod. The perturbed quantities are preceded by $\delta$. Carrying out the first variation of Eq.~(\ref{eqbendingenergyquaternions}), the corresponding variation of the energy $\mathcal{E}_{e}$ is,
\begin{equation}  
\delta\mathcal{E}_{e} = \sum_{k=1}^3 E_kI_k\int_0^L\left(2\mat{B}_k\vect{q}\vect{q}' - \hat{\kappa}_k\right) \left(2\mat{B}_k\vect{\delta q}\vect{q}' + 2\mat{B}_k\vect{q}\vect{\delta q}'\right)ds,
\label{eqdeltabendingenergy1}
\end{equation}
where $\vect{\delta q} =  \tp{\left[\delta q_1 \, \delta q_2 \, \delta q_3 \, \delta q_4\right]}$ is the vector of the arbitrary perturbations of the rotational degrees of freedom $\vect{q}$. Upon integration by parts, we transform Eq.~(\ref{eqdeltabendingenergy1}) into an integral that depends on $\vect{\delta q}$ alone to arrive at,
\begin{multline} 
\delta\mathcal{E}_{e} = \left[ \sum_{k=1}^3 E_kI_k\left(2\mat{B}_k\vect{q}\vect{q}' - \hat{\kappa}_k\right)2\mat{B}_k\vect{q}\vect{\delta q}\right]^L_0 - \int_0^L \sum_{k=1}^3 E_kI_k ... \\
\left[\left(2\mat{B}_k\vect{q}\vect{q}' - \hat{\kappa}_k\right)4\mat{B}_k\vect{q}' + \left(2\mat{B}_k\vect{q}'\vect{q}' + 2\mat{B}_k\vect{q}\vect{q}'' - \hat{\kappa}_k'\right)2\mat{B}_k\vect{q}\right] \vect{\delta q} ds,
\label{eqdeltabendingenergy2}
\end{multline}
where the first term stands for the variation of elastic energy over the entire interval and is the boundary term from the integration by parts assuming that the rod is parametrized from $s=0$ to $s=L$. Physically, this first term represents the work done by the operator upon a change of orientation applied to the ends of the rod. We can rewrite this term in the concise form $[\vect{T}(s) \vect{\delta q}]$ where, 
\begin{equation}  
\vect{T}(s) = G_1(s) 2\mat{B}_1\vect{q} + G_2(s) 2\mat{B}_2\vect{q} + G_3(s) 2\mat{B}_3\vect{q}.
\label{eqinternalmoment}
\end{equation}
The vector $\vect{T}(s)  =  \tp{\left[T_1(s) \, T_2(s) \, T_3(s) \, T_4(s)\right]}$ is the internal moment projected in the quaternion basis defined 
as a linear superposition of the internal moments due to elementary modes of deformation. The functionals $G_1(s)$, $G_2(s)$ and $G_3(s)$ given by,
\begin{equation}  
G_k(s) = E_kI_k \left(\kappa_k(s) - \hat{\kappa}_k(s)\right)  = E_kI_k \left(2\mat{B}_k\vect{q}(s),\vect{q}'(s) - \hat{\kappa}_k(s)\right) 
\label{eqinternalcouples}
\end{equation}
are respectively the two flexural and torsional moments, defined as the components of $\vect{T}(s)$ in the local material frame. The second term in Eq.~(\ref{eqdeltabendingenergy2}) is the work done by the operator upon a change of orientation applied along the rod. The elementary contribution to the integral can be rewritten $\int_0^L \vect{\tau}(s) \vect{\delta q} ds$ where $\vect{\tau}(s)  =  \tp{\left[\tau_1(s) \, \tau_2(s) \, \tau_3(s) \, \tau_4(s)\right]}$ as a four-dimensional vector written in the quaternion basis that reads,
\begin{equation}  
\vect{\tau}(s) = \sum_{k=1}^3\left(G_k(s) 4\mat{B}_k\vect{q}'(s) + G'_k(s) 2\mat{B}_k\vect{q}(s)\right),
\label{eqnetinternalmoment}
\end{equation}
where $G'_k(s) = E_kI_k \left(2\mat{B}_k\vect{q}'(s)\vect{q}'(s) + 2\mat{B}_k\vect{q}(s)\vect{q}''(s) - \hat{\kappa}_k'(s)\right)$ is the differential of $G_k(s)$ with respect to $s$. The quantity $\vect{\tau}(s) ds$ is the net moment applied on an infinitesimal element of the rod located between the cross-sections at $s$ and $s+ds$.

Before arriving to the equilibrium equations from this variation, we need to consider the external loads that are applied to the rod, and whose work must balance the variation of energy at equilibrium. Here, we consider two types of external loads: point forces $\left(\vect{P}(0),\vect{P}(L)\right)$ and torques $\left(\vect{M}(0),\vect{M}(L)\right)$ that are applied at the two ends $s=0$ and $s=L$, and distributed forces and torques that are applied along the length of the rod, with linear densities $\vect{p}(s)$ and $\vect{m}(s)$, respectively. The density of forces, $\vect{p}(s)$, can represent, for instance, the weight of the rod, and the density of moments, $\vect{m}(s)$, hydrostatic loadings such as the result of viscous stresses due to a swirling flow around the rod. The total work done by these external forces upon an infinitesimal perturbation of the rod's configuration is,
\begin{multline}  
\delta\mathcal{W} = \vect{P}(0)\vect{\delta r}(0) + \vect{M}(0)\vect{\delta q}(0) + \vect{P}(L)\vect{\delta r}(L) + \vect{M}(L)\vect{\delta q}(L) + \\
                                   \int_0^L \left(\vect{p}(s) \vect{\delta r}(s) + \vect{m}(s) \vect{\delta q}(s)\right) ds,
\label{eqexternalwork}
\end{multline}
where $\vect{\delta r} =  \tp{\left[\delta r_1 \, \delta r_2 \, \delta r_3\right]}$ is the vector of the small arbitrary perturbations of the translational degrees of freedom $\vect{r}$. According to Eq.~(\ref{eqexternalwork}), the external forces $\vect{P}(s) = \tp{\left[P_x(s) \, P_y(s) \, P_z(s)\right]}$ and $\vect{p}(s) = \tp{\left[p_x(s) \, p_y(s) \, p_z(s)\right]}$ are defined in terms of the global directions $x$, $y$, $z$ whereas the external moments $\vect{M}(s) = \tp{\left[M_1(s) \, M_2(s) \, M_3(s) \, M_4(s) \right]}$ and $\vect{m}(s) = \tp{\left[m_1(s) \, m_2(s) \, m_3(s) \, m_4(s) \right]}$ are expressed in the quaternion basis $i$, $j$, $k$, $1$ of $\mathcal{H}$, defined by Eq.~(\ref{eqformulaquaternions}). 
In Section \ref{sec5}, we will show through the specific example of the writhing of a rod how to express physical rotational quantities (e.g. boundary conditions or external moments) in terms of quaternions.


\subsection{Equilibrium equations}

Thus far, we have implicitly assumed that the perturbations $\tp{\left[\vect{\delta r} \, \vect{\delta q}\right]}$ can be chosen freely. This is, however, not the case since our rod is subject to the kinematical constraints  introduced previously in Eqs.~(\ref{equnitquaternion}) and (\ref{eqadaptedframeconstraint}). These constraints are imposed in the derivation of the equations of equilibrium by adding a number of Lagrange multipliers into the variation of the elastic energy $\mathcal{E}_{e}$ and external loads $\delta\mathcal{W} $. In this Lagrangian formalism, the enforcement of the unicity of quaternion in Eq.~(\ref{equnitquaternion}), translates as the continuous functional constraint,
\begin{equation}  
\mathcal{C}_{\alpha}\left[\vect{q}(s)\right] = \vect{q}(s)\vect{q}(s) - 1 = 0,
\label{eqconstraintquaternion}
\end{equation}
where the brackets indicate that $\mathcal{C}_{\alpha}$ depends on the function $\vect{q}(s)$, globally. Moreover, Eq.~(\ref{eqadaptedframeconstraint}), which ensures that the directors are adapted to the Cosserat curve, translates to three conditions on the continuous vector-valued function,    
\begin{equation}  
\vect{\mathcal{C}}_{\vect{\mu}}\left[\vect{r}(s), \vect{q}(s)\right] = \vect{r}'(s) - \vect{d}_3(\vect{q}(s)) = \vect{0}.
\label{eqconstraintadapted}
\end{equation} 

With the expressions for the energy of an arbitrary configuration of the rod in Eqs.~(\ref{eqdeltabendingenergy2}) and (\ref{eqexternalwork}) in hand, the equations of equilibrium are now obtained by assuming that the energy is stationary under small deformations for a given set of  boundary conditions and geometrical constraints; Eqs.~(\ref{equnitquaternion}) and (\ref{eqadaptedframeconstraint}). This is equivalent to requiring that the first order variation of the functionals $\delta\mathcal{E}_{e}$ and $\delta\mathcal{W}$, combined linearly with the variation of the constraints $\delta \mathcal{C}_{\alpha}$ and $\vect{\delta \mathcal{C}}_{\vect{\mu}}$ over the interval from $s=0$ to $s=L$ (i.e. the Lagrangian) vanish,
\begin{equation}  
\delta\mathcal{E}_{Lag} = \delta\mathcal{E}_{e} - \delta\mathcal{W} +  \int_0^L \alpha(s) \delta\mathcal{C}_{\alpha}(s) ds + \int_0^L \vect{\mu}(s)\vect{\delta \mathcal{C}}_{\vect{\mu}}(s) ds = 0.
\label{eqvariationprinciple}
\end{equation}
In this equation, the variation of the constraint $\mathcal{C}_{\alpha}(s)$ given in Eq.~(\ref{eqconstraintquaternion}) takes the form,
\begin{equation}  
\int_0^L \alpha(s) \delta\mathcal{C}_{\alpha}(s) ds = \int_0^L 2\alpha \vect{q} \vect{\delta q} ds,
\label{eqvariationconstraintquaternion}
\end{equation}
where the scalar function $\alpha(s)$ is the Lagrange multiplier that imposes the norm of the quaternions to be one. The variation of the constraints $\vect{\mathcal{C}}_{\vect{\mu}(s)}$ given in Eq.~(\ref{eqconstraintadapted}) reads, after integration by parts,
\begin{equation}  
\int_0^L \vect{\mu}(s)\vect{\delta \mathcal{C}}_{\vect{\mu}}(s) ds = \left[\vect{\mu} \vect{\delta r}\right]^L_0 - \int_0^L\vect{\mu}' \vect{\delta r} + 2\mat{D}(\vect{q})\vect{\mu}\vect{\delta q} ds,
\label{eqvariationconstraintadapted}
\end{equation}
where the terms of the vector valued function $\vect{\mu}(s) =  \tp{\left[\mu_{x}(s) \,\mu_{y}(s) \, \mu_{z}(s)\right]}$ are the Lagrange multipliers ensuring the condition of inextensibility of the slender elastic rods and the operator $\mat{D}(\vect{q})$ reads,
\begin{equation}  
\mat{D}(\vect{q}(s)) = \left[ \begin{array}{ccc}
q_3(s) & -q_4(s) & -q_1(s) \\
q_4(s) & q_3(s) & -q_2(s) \\
q_1(s) & q_2(s) & -q_3(s)\\
q_2(s) & -q_1(s) & -q_4(s)
\end{array} \right].
\label{eqderived3}
\end{equation}
Now, substituting Eqs.~(\ref{eqdeltabendingenergy2}), (\ref{eqexternalwork}), (\ref{eqvariationconstraintquaternion}) and (\ref{eqvariationconstraintadapted}) into the Lagrangian of Eq.~(\ref{eqvariationprinciple}), we arrive at the first variation of the geometrically constraint elastic energy of the slender elastic rod,
\begin{multline}  
\delta\mathcal{E}_{Lag} =  \left[\vect{T}(s) \vect{\delta q}(s) + \vect{\mu}(s)\vect{\delta r}(s)\right]^L_0 - \vect{M}(0)\vect{\delta q}(0) - \vect{M}(L)\vect{\delta q}(L) \\
                                               - \vect{P}(0)\vect{\delta r}(0) - \vect{P}(L)\vect{\delta r}(L)  - \int_0^L  \left(\vect{p}(s) + \vect{\mu}'(s) \right)\vect{\delta r}(s) ds \\
                                               - \int_0^L \left(\vect{\tau}(s) + \vect{m}(s) - 2\alpha(s) \vect{q}(s) + 2\mat{D}(\vect{q}(s))\vect{\mu}(s)\right) \vect{\delta q}(s) ds.
\label{eqlagrangian}
\end{multline}
The condition that the variation in Eq.~(\ref{eqlagrangian}) must vanish for an arbitrary perturbations $\vect{\delta r}(s)$ and $\vect{\delta q}(s)$ yields the strong form of the equilibrium equations for our elastic rod as second-order differential equations,
\begin{subequations}
\label{eqequilibriumequations}
\begin{align}
\vect{0} & = \vect{p}(s) + \vect{\mu}'(s) \label{eqequilibriumequationsa} \\
\vect{0} & = \vect{\tau}(s) - \vect{m}(s) + 2\alpha(s) \vect{q}(s) - 2\mat{D}(\vect{q}(s))\vect{\mu}(s). \label{eqequilibriumequationsb}
\end{align}  
\end{subequations}
When projected along the three directions of the global cartesian frame $\left(x,y,z\right)$, the vector equation Eq.~(\ref{eqequilibriumequationsa}) yields a set of three differential equations that can be interpreted as the balance of forces. The vector of Lagrange multiplier $\vect{\mu}(s)$ measures the resultant of the contact forces transmitted through the rod's cross-section. Indeed, calculating the forces acting on a small element of the rod of length $ds$, we find that the element is submitted to the contact forces $\vect{\mu}(s + ds)$ and $-\vect{\mu}(s)$ from the neighboring elements, and to the external force $\vect{p}ds$. At equilibrium, the total forces $\left(\vect{\mu}'ds + \vect{p}ds\right)$ is zero as described by Eq.~(\ref{eqequilibriumequationsa}). 

When projected along the four elements $\left(i, j, k, 1\right)$ of the quaternion basis $\mathcal{H}$, the vector equation Eq.~(\ref{eqequilibriumequationsb}) yields a set of four differential equations that can be interpreted as the balance of moments. Working in the quaternion basis, it is, however, not straightforward  to find an obvious physical interpretation for each of the terms but it suffices to say that they are related to the internal moments acting on a small element of the rod $ds$.

For the equilibrium equations written in Eqs.~(\ref{eqequilibriumequations}) to be complete and well-posed, one must add the geometrical constraints given by Eq.~(\ref{eqconstraintquaternion}) and Eqs.~(\ref{eqconstraintadapted}), which, in their projected and developed form, read as,
\begin{subequations}
\label{eqconstraintsprojected}
\begin{align}
0 & = q_1^2(s) + q_2^2(s) + q_3^2(s) + q_4^2(s) - 1 \label{eqconstraintsprojecteda} \\
0 & = r'_x(s) - 2q_1(s)q_3(s) - 2q_2(s)q_4(s) \label{eqconstraintsprojectedb} \\
0 & = r'_y(s) - 2q_2(s)q_3(s) + 2q_1(s)q_4(s)  \label{eqconstraintsprojectedc} \\
0 & = r'_z(s) + q_1^2(s) + q_2^2(s) - q_3^2(s) - q_4^2(s). \label{eqconstraintsprojectedd}
\end{align}  
\end{subequations} 
In the seven differential equilibrium equations of Eqs.~(\ref{eqequilibriumequations}) plus the four differential equations in Eqs.~(\ref{eqconstraintsprojected}), the eleven unknowns  are the four Lagrange multipliers $\alpha(s)$, $\mu_{x}(s)$, $\mu_{y}(s)$ and $\mu_{z}(s)$, the four rotational degrees of freedom $q_1(s)$, $q_2(s)$, $q_3(s)$ and $q_4(s)$ and the three translational degrees of freedom $r_x(s)$, $r_y(s)$ and $r_z(s)$. 
Thanks to the use of quaternions, the kinematics is geometrically-exact and the resultant equilibrium equations are simply polynomial since the highest geometric nonlinearity comes from the vector $\vect{\tau}(s)$ given in Eq.~(\ref{eqinternalmoment}), which is cubic in $\vect{q}(s)$. In Section \ref{sec4}, while developing the numerical implementation, we will make extensive use of this smooth and regular nonlinearity to efficiently compute the numerical solutions of these equations. 

So far, in Eq.~(\ref{eqlagrangian}), we have only considered the vanishing of the integral term. Likewise, boundary terms should also vanish since this equation is also to be satisfied for perturbations localized at its extremities. The first boundary terms, associated with rotations $\vect{\delta q}(L)$ and $\vect{\delta q}(0)$ yield,
\begin{subequations}
\label{eqboundaryquaternions}
\begin{align}
0 & = \left(\vect{T}(0) + \vect{M}(0)\right)\vect{\delta q}(0), \label{eqboundaryquaternionsa} \\
0 & = \left(\vect{T}(L) - \vect{M}(L)\right)\vect{\delta q}(L).  \label{eqboundaryquaternionsb}
\end{align}  
The remaining boundary terms associated with displacements $\vect{\delta r}(0)$ and $\vect{\delta r}(L)$ of the ends $s = 0$ and $s=L$, respectively, yield,
\begin{align}
0 & = \left(\vect{\mu}(0) + \vect{P}(0)\right) \vect{\delta r}(0), \label{eqboundaryquaternionsc} \\
0 & = \left(\vect{\mu}(L) - \vect{P}(L)\right) \vect{\delta r}(L). \label{eqboundaryquaternionsd}
\end{align}  
\end{subequations} 
To provide a physical interpretation of the behavior at the boundary conditions, we first consider Eq.~(\ref{eqboundaryquaternionsa}). If the endpoint $s = 0$ is free to rotate, the vector $\vect{\delta q}(0)$ is arbitrary and one is led to the boundary condition $\vect{T}(0) + \vect{M}(0) = \vect{0}$. This is the total torque applied on the section $s=0$, which is the sum of the internal moments $\vect{T}(0)$ transmitted by the downstream part of the rod, $s > 0$, and of the moment $\vect{M}(0)$ applied by the operator. At equilibrium, the total torque should vanish when the end is free to rotate. If the endpoint $s = 0$ is fixed, the perturbations that are consistent with the kinematics are such that $\vect{\delta q}(0) = \vect{0}$ and the equation is automatically satisfied. The boundary condition is then the one imposing the rotation of the fixed end, which leaves the total number of boundary conditions unchanged. The same reasoning holds for Eq.~(\ref{eqboundaryquaternionsb}) near the opposite end, $s = L$, although the total torque is now $\vect{T}(L) - \vect{M}(L)$, since, in this case the internal moment is applied by the downstream part of the rod, $s<L$.

The two other boundary conditions written in Eqs.~(\ref{eqboundaryquaternionsc})-(\ref{eqboundaryquaternionsd}) can be handled in a similar fashion. Near an end where the displacement is unconstrained, the total force should be zero. This total force is  $\vect{\mu}(0) + \vect{P}(0)$ near the end $s=0$, by a similar reasoning as above. However, the total force is $\vect{\mu}(L) - \vect{P}(L)$ near the opposite end, $s=L$, given that the internal forces $\vect{\mu}(L)$ are now applied by the downstream part of the rod, $s<L$. This remark validates our previous interpretation as for the physical interpretation of the Lagrange multipliers $\mu_{x}(s)$, $\mu_{y}(s)$ and $\mu_{z}(s)$; they are the internal forces along the three directions of the global frame that constrain the directors to be adapted to the Cosserat curve.  

Together, Eqs.~(\ref{eqequilibriumequations})-(\ref{eqboundaryquaternions}) constitute the set of geometrically-exact cubic differential equations that describe the mechanical behavior of the slender elastic rod represented in Fig.~\ref{fig:Cosserat_rod}(a). These nonlinear differential equations could be solved with classic boundary value problem algorithms upon knowing the boundary conditions in terms of external forces or kinematics. Moreover, coupled with traditional predictor-corrector methods, one should be able to continue, step-by-step, the solutions of this nonlinear elastic problem in terms of given geometric or mechanical control parameters \cite{crisfield1991nonlinear,doedel1981auto}. In the following section, in an alternative point of departure, we use a continuation method based on the Asymptotic Numerical Method (ANM) developed in the early 1990's to solve elastic structural problems in the early post-buckled regime \cite{damil1990new,cochelin1994path}. Taking advantage of the particular cubic form of the geometrically-exact equilibrium Eqs~(\ref{eqequilibriumequations})-(\ref{eqboundaryquaternions}), this path-following perturbation technique will enable the determination of semi-analytical nonlinear solution branches by inverting a simple stiffness matrix at each step of the continuation. This outstanding numerical property makes the ANM algorithm highly robust and computationally efficient at determining the various equilibria of our slender elastic rod.

\section{Numerical method}\label{sec4}

In this Section, we solve the differential equilibrium equations Eqs.~(\ref{eqequilibriumequations})-(\ref{eqboundaryquaternions}) using a finite element-based semi-analytical path-following method. We first approximate the continuous degrees of freedom using finite differences approximation to interpolate the mechanical and geometrical variables at each nodes and elements. Thanks to the quaternion formalism introduced above, the equilibrium equations can be reduced to an algebraic set of quadratic equations by considering the flexural and torsional internal moments as unknowns. This quadratic form is particularly well suited to ANM which is a semi-analytical continuation algorithm to compute the branches of solution of a set of nonlinear polynomial equations. To follow all the bifurcated branches, we show how the local stability of the computed equilibria can be assessed by using the second order conditions of constrained minimization problems.
Finally, we describe the implementation of our algorithm into MANlab, a free and interactive bifurcation analysis software based in MATLAB.

\subsection{Discretization}\label{discretization}

In order to compute the equilibrium equations, Eqs~(\ref{eqequilibriumequations})-(\ref{eqboundaryquaternions}), we first explain how to discretize the main function unknowns such as strain rate vector $\vect{\kappa}(s)$, material frame $\left(\vect{d}_1(s) \, \vect{d}_2(s) \, \vect{d}_3(s)\right)$, positional and rotational degrees of freedom $\vect{r}(s)$ and $\vect{q}(s)$ or Lagrange multipliers $\alpha(s)$ and $\vect{\mu}(s)$. 

\begin{figure}[t!]
\includegraphics[width=\columnwidth]{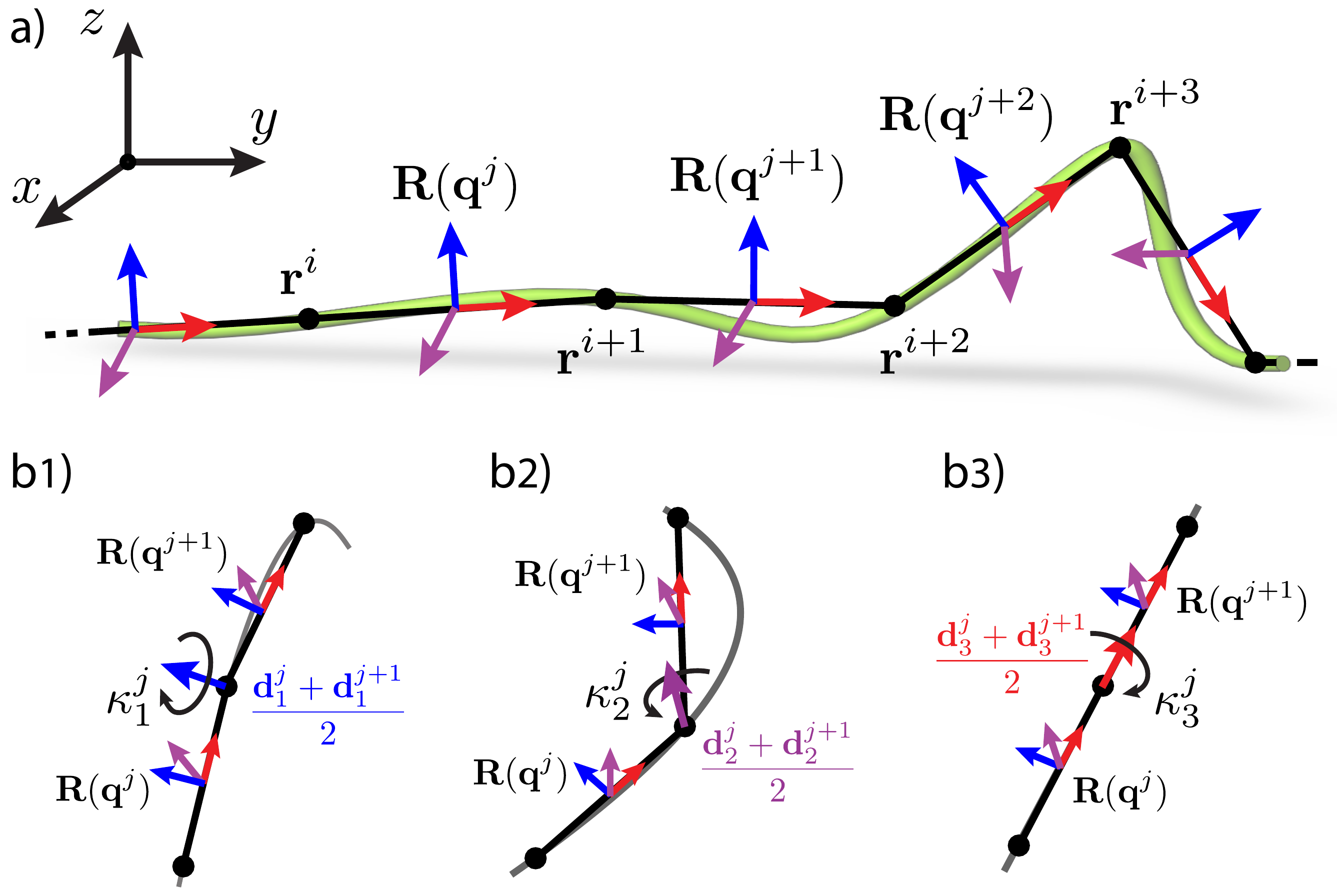}
\caption{\label{fig:DIscrete_Cosserat_rod} Finite element discretization method. (a) The centerline of the rod is discretized into $N$ elements separated by $N+1$ nodes, $\vect{r}^j$. We also consider $N$ material frames $\mat{R}(\vect{q}^j)$ to express the orientation of the $j^{\text{th}}$ element. (b) The change of orientation between two successive elements $j$ and $j+1$ is expressed by the $N-1$ discrete material curvatures at the interconnected nodes: (b1) $\kappa_1^j$ around the director $(\vect{d}_1^j+\vect{d}_1^{j+1})/2$, (b2) $\kappa_2^j$ around the director $(\vect{d}_2^j+\vect{d}_2^{j+1})/2$, (b3) $\kappa_3^j$ around the director $(\vect{d}_3^j+\vect{d}_3^{j+1})/2$.}
\end{figure}

The position of the rod is represented by discretizing its centerline into $N$ elements separated by $N+1$  spatial control points, $\vect{r}(s^i) = \vect{r}^i = \tp{\left[r_x^i \, r_y^i \, r_z^i\right]}$ in $\mathcal{R}^3$, located by the discrete curvilinear coordinate $s^i$ as illustrated in Fig.~\ref{fig:DIscrete_Cosserat_rod}(a). The spatial derivative of the positional degrees of freedom is approximated by the forward finite difference between two successive nodes, 
\begin{equation}  
\vect{r}'(s^i) \approx \frac{\vect{r}(s^i+ds^i) - \vect{r}(s^i)}{ds^i} = \frac{\vect{r}^{i+1} - \vect{r}^i}{ds^i}
\label{eqderivativeposition}
\end{equation}
where $ds^i = \|\vect{r}^{i+1} - \vect{r}^i\| = L/N$ is the length of the $i^{\text{th}}$ element and $L$ is the total length of the rod. To ensure the inextensibility condition of our rods, $ds^i$ is constant upon deformation and the stretch along the centerline is forced to verify $\|\vect{r}'(s^i)\| = 1$. 

The orientations of the centerline elements are represented by employing $N$ material frames $\mat{R}(\vect{q}^j) = \left[\vect{d}_1^j \, \vect{d}_2^j \, \vect{d}_3^j\right]$ in $\mathcal{R}^{3\times 3}$ where $\vect{q}^j =  \tp{\left[q_1^j \, q_2^j \, q_3^j \, q_4^j\right]} $ is the set of quaternions associated with each element $j$. According to Eq.~(\ref{eqrotationmatrixquaternion}), the directors of the $j^{\text{th}}$ element are vectors in $\mathcal{R}^3$ represented at the midpoints on the centerline segments [see Fig.~\ref{fig:DIscrete_Cosserat_rod}(a)] such that,
\begin{gather}  
\vect{d}_1^j = \left[\begin{array}{lll}
q_1^jq_1^j - q_2^jq_2^j - q_3^jq_3^j  + q_4^jq_4^j  \\
2(q_1^jq_2^j + q_3^jq_4^j) \\
2(q_1^jq_3^j - q_2^jq_4^j) \end{array} \right], \quad \vect{d}_2^j  = \left[ \begin{array}{lll}
2(q_1^jq_2^j - q_3^jq_4^j) \\
-q_1^jq_1^j + q_2^jq_2^j - q_3^jq_3^j + q_4^jq_4^j \\
 2(q_2^jq_3^j + q_1^jq_4^j) \end{array} \right], \nonumber\\
\vect{d}_3^j = \left[ \begin{array}{lll}
2(q_1^jq_3^j + q_2^jq_4^j) \\
 2(q_2^jq_3^j - q_1^jq_4^j) \\
-q_1^jq_1^j - q_2^jq_2^j + q_3^jq_3^j + q_4^jq_4^j \end{array} \right]. 
\label{eqrotationmatrixdiscrete}
\end{gather}
Replacing the quaternions functions $\vect{q}(s)$ by their discrete counterparts $\vect{q}^j$ in the expression of strain rates given in Eq.~(\ref{eqstrainratesfinal}), we can write the discrete material curvatures $\kappa_1^j$, $\kappa_2^j$ and the twist $\kappa_3^j$  expressing the extent of rotation around the directors, $\vect{d}_1^j$, $\vect{d}_2^j$ and $\vect{d}_3^j$, between two successive elements [see Fig.~\ref{fig:DIscrete_Cosserat_rod}(b)] in the form
\begin{equation}  
\kappa_k^j = 2\mat{B}_k\bar{\vect{q}}^j\vect{q}'^j \quad \textnormal{for k=}1, 2, 3.
\label{eqstraindiscrete}
\end{equation}
In Eq.~(\ref{eqstraindiscrete}), we introduced the average and the spatial derivative of the rotational degrees of freedom of the $j^{\text{th}}$ element $\vect{q}^j$ as,
\begin{equation}  
\bar{\vect{q}}^j = \frac{\vect{q}^{j+1} + \vect{q}^j}{2} \quad \text{and} \quad \vect{q}'^j = \frac{\vect{q}^{j+1} - \vect{q}^j}{ds^j},
\label{eqderivativequaternion}
\end{equation}
where $ds^j = 1/2(\|\vect{r}^{i+2} - \vect{r}^{i+1}\|+ \|\vect{r}^{i+2} - \vect{r}^{i+1}\|) = L/N$, taking into consideration the rod's inextensibility condition.

In a similar fashion, replacing the continuous function $\vect{q}(s)$ and its derivative $\vect{q}'(s)$ by their discretized counterparts, $\bar{\vect{q}}^j$ and $\vect{q}'^j$,  respectively, given in Eq.~(\ref{eqderivativequaternion}), the first variation of elastic energy previously given in Eq.~(\ref{eqdeltabendingenergy1}) can be approximated by a Riemann sum over the elements from $j = 1$ to $N$,
\begin{equation}  
\delta\mathcal{E}_{e} \approx \sum_{k=1}^3 \sum_{j=1}^{N-1} 2G_k^j \left(\mat{B}_k\vect{q}^j\vect{\delta q}^{j+1} - \mat{B}_k\vect{q}^{j+1}\vect{\delta q}^j\right)ds^j.
\label{eqdeltabendingenergy1discrete}
\end{equation}
In this equation, the $N$ vectors $\vect{\delta q}^j = \tp{\left[\delta q_1^j \, \delta q_2^j \, \delta q_3^j \, \delta q_4^j\right]} $ are the discrete version of the perturbed rotational degrees of freedom $\vect{\delta q}(s)$ and are associated with each element $j$. The $3(N-1)$ constants $G_k^j$ are an approximation of the flexural and torsional internal torques $G_k(s)$ given in Eq.~(\ref{eqinternalcouples}), which read
\begin{equation}  
G_k^j = E_kI_k \left(\mat{B}_k\left(\vect{q}^j+\vect{q}^{j+1}\right)\frac{1}{ds^j}\left(\vect{q}^{j+1}-\vect{q}^j\right) - \hat{\kappa}_k^j\right),
\label{eqinternalmomentcoeffdiscrete}
\end{equation}
and are defined between two successive elements for $1 \leq j \leq N$.

Replacing $\vect{q}(s)$ and $\vect{r}(s)$ by their discrete counterparts at each element $\vect{q}^j$ and each node $\vect{r}^i$, respectively, the variation of the work done by the external forces and torques previously given in Eq.~(\ref{eqexternalwork}), can be approximate by its discrete version,
\begin{multline}  
\delta\mathcal{W} \approx \vect{P}^0\vect{\delta r}^1 + \vect{M}^0\vect{\delta q}^1 + \vect{P}^L\vect{\delta r}^{N+1} + \vect{M^L}\vect{\delta q}^N + \\
                                   \sum_{i=2}^{N} \vect{p}^i \vect{\delta r}^ids^i + \sum_{j=2}^{N-1}\vect{m}^j \vect{\delta q}^j ds^j,
\label{eqexternalworkdiscrete}
\end{multline}
where $\vect{\delta r}^i = \tp{\left[\delta r_x^i \, \delta r_y^i \, \delta r_z^i\right]}$ is the vector of the perturbed displacement at each node. In this equation, we have introduced the vector of point forces at the two ends $\vect{P}^0 = \tp{\left[P_x^0 \, P_y^0 \, P_z^0\right]}$ and $\vect{P}^L = \tp{\left[P_x^L \, P_y^L \, P_z^L\right]}$ and the vector of density of external forces at each node $\vect{p}^i = \tp{\left[p_x^i \, p_y^i \, p_z^i\right]}$ for $2 \leq i \leq N$ defined in term of the global directions $x$, $y$, $z$. In the discrete version of $\delta\mathcal{W}$, Eq.~(\ref{eqexternalworkdiscrete}), we have also introduced the torques applied at the two ends $\vect{M}^0 = \tp{\left[M_1^0 \, M_2^0 \, M_3^0 \, M_4^0\right]}$ and $\vect{M}^L = \tp{\left[M_1^L \, M_2^L \, M_3^L \, M_4^L\right]}$ and the vector of density of external moment at each element $\vect{m}^j = \tp{\left[m_1^j \, m_2^j \, m_3^j \, m_4^j\right]}$ for $2 \leq j \leq N-1$, also expressed in the quaternion basis. 

Before we derive the algebraic system of equilibrium equations, we still need to write the discrete form of the variation of work due to the geometrical constraints in Eqs.~(\ref{eqconstraintquaternion}) and~(\ref{eqconstraintadapted}). Replacing $\vect{q}(s)$ by its discrete counterpart, $\vect{q}^j$, we can expand Eq.~(\ref{eqvariationconstraintquaternion}) in the form of a Riemann sum,
\begin{equation}  
\int_0^L \alpha(s) \delta\mathcal{C}_{\alpha}(s) ds \approx \sum_{j=1}^{N} 2\alpha^j \vect{q}^j \vect{\delta q}^j ds^j,
\label{eqvariationconstraintquaterniondiscrete}
\end{equation}
where $\alpha^j$ is a discrete scalar at each element $j$, approximating the continuous Lagrange parameter given in Eq.~(\ref{eqvariationconstraintquaternion}). 
Introducing the $N$ vectors of Lagrange parameters $\vect{\mu}^j =  \tp{\left[\mu_{x}^j \,\mu_{y}^j \, \mu_{z}^j\right]}$ which prescribe that each element is parallel to the tangent $\vect{r}'(s^i)$ given in (\ref{eqderivativeposition}), we can rewrite Eq.~(\ref{eqvariationconstraintadapted}) in its discrete form,
\begin{multline}  
\int_0^L \vect{\mu}(s)\vect{\delta \mathcal{C}}_{\vect{\mu}}(s) ds \approx \vect{\mu}^N \vect{\delta r}^{N+1}  - \vect{\mu}^1\vect{\delta r}^{1}  - \\
                                                                                                       \sum_{i=2}^{N} \left(\vect{\mu}^i - \vect{\mu}^{i-1}\right)\vect{\delta r}^i 
                                                                                                       + \sum_{j=1}^{N}\mat{D}(\vect{q}^j)\vect{\mu}^j\vect{\delta q}^j ds^j,
\label{eqvariationconstraintadapteddiscrete}
\end{multline}
where the vector of Lagrange parameters, or internal contact forces, $\vect{\mu}(s)$, has been approximated by the backward finite difference between each successive elements,
\begin{equation}  
\vect{\mu}'^j  = \frac{\vect{\mu}^{j} - \vect{\mu}^{j-1}}{ds^j},
\label{eqderivativelagrange}
\end{equation}
and $\mat{D}(\vect{q}^j)$ is the discrete counterpart of $\mat{D}(\vect{q}(s))$ introduced in Eq.~(\ref{eqderived3}) which reads, at each element,
\begin{equation}  
\mat{D}(\vect{q}^j) = \left[ \begin{array}{ccc}
q_3^j & -q_4^j & -q_1^j \\
q_4^j & q_3^j & -q_2^j \\
q_1^j & q_2^j & -q_3^j\\
q_2^j & -q_1^j & -q_4^j
\end{array} \right].
\label{eqderived3discrete}
\end{equation}

As we did previously in the continuum case, we now require that the discrete variation of the Lagrangian $\delta\mathcal{E}_{Lag}$ (given in Eq.~(\ref{eqvariationprinciple}) as the sum of Eqs.~(\ref{eqdeltabendingenergy1discrete}), (\ref{eqexternalworkdiscrete}), (\ref{eqvariationconstraintquaterniondiscrete}) and (\ref{eqvariationconstraintadapteddiscrete})) vanishes for any arbitrary perturbations $\vect{\delta r}^i$ and $\vect{\delta q}^j$. This condition yields the set of algebraic equilibrium equations of the discrete unshearable and inextensible slender elastic rod.  The condition that this variation is zero for any perturbed displacements $\vect{\delta r}^i$ leads to the balance of forces as a set of algebraic equations,
\begin{subequations}
\label{eqequilibriumforcediscrete}
\begin{align}
\vect{P}^0 + \vect{\mu}^i = \vect{0} & \quad \text{for $ i = 1$} \label{eqequilibriumforcediscretea} \\
\vect{p}^iL/N + \vect{\mu}^{i} - \vect{\mu}^{i-1} = \vect{0} & \quad \text{for $2 \leq i \leq N$} \label{eqequilibriumforcediscreteb} \\
\vect{P}^L - \vect{\mu}^{i-1} = \vect{0} & \quad \text{for $i = N+1$}. \label{eqequilibriumforcediscretec} 
\end{align}  
\end{subequations} 
Projected along the three directions of the global cartesian frame $(x, y, z)$, Eqs.~(\ref{eqequilibriumforcediscrete}) yield $3(N+1)$ linear equations for the $3N$ unknowns $\vect{\mu}^j$. In the limit $N$ very large, these equations converge to the continuous differential equations Eq.~(\ref{eqequilibriumequationsa}). The condition that the variation of the Lagrangian is zero for any arbitrary perturbations $\vect{\delta q}^j$ leads to the balance of moments as a set of discrete algebraic equations,
\begin{subequations}
\label{eqequilibriummomentdiscrete}
\begin{align}
\vect{\tau}^j  + \vect{M}^0 + 2\mat{D}(\vect{q}^j)\vect{\mu}^j - 2\alpha^j\vect{q}^j = \vect{0} & \quad \text{for $ j = 1$} \label{eqequilibriummomentdiscretea} \\
\vect{\tau}^j + \vect{m}^j + 2\mat{D}(\vect{q}^j)\vect{\mu}^j - 2\alpha^j\vect{q}^j = \vect{0}  & \quad \text{for $2 \leq j \leq N-1$} \label{eqequilibriummomentdiscreteb} \\
\vect{\tau}^j  - \vect{M}^L + 2\mat{D}(\vect{q}^j)\vect{\mu}^j - 2\alpha^j\vect{q}^j  = \vect{0} & \quad \text{for $j = N$}. \label{eqequilibriummomentdiscretec} 
\end{align}  
In these equations, $\vect{\tau}^j =  \tp{\left[\tau_1^j \, \tau_2^j \, \tau_3^j \, \tau_4^j\right]}$ is the vector of net internal moment applied on the element $j$ written in the quaternion basis,
\begin{align}
\vect{\tau}^j  & = \sum\nolimits_{k=1}^3 G^j_k 2\mat{B}_k \vect{q}^{j+1}  &  \text{for $ j = 1$} \label{eqequilibriummomentdiscreted} \\
\vect{\tau}^j & =  \sum\nolimits_{k=1}^3 \left(G^{j-1}_k 2\mat{B}_k \vect{q}^{j-1} - G^{j}_k 2\mat{B}_k \vect{q}^{j+1} \right)  & \text{for $2 \leq j \leq N-1$} \label{eqequilibriummomentdiscretee} \\
\vect{\tau}^j  & =  \sum\nolimits_{k=1}^3 G^{j-1}_k 2\mat{B}_k \vect{q}^{j-1} &  \text{for $j = N$}.\label{eqequilibriummomentdiscretef}
\end{align}  
\end{subequations} 
and $2\mat{D}(\vect{q}^j)\vect{\mu}^j$ is the moment resultant from the internal contact forces $\vect{\mu}^j$ applied on the element $j$. In the limit of large $N$,  Eqs.~(\ref{eqequilibriummomentdiscrete}) converge to the continuous differential equation Eq.~(\ref{eqequilibriumequationsa}). Projected along the four elements $(i,j,k,1)$ of the quaternion basis $\mathcal{H}$, Eqs.~(\ref{eqequilibriummomentdiscrete}) yield $4N$ nonlinear equations for the $4N$ unknowns $\vect{q}^j$ and $N$ unknowns $\alpha^j$. The missing equations required to compute all the unknowns are given by the geometrical constraints Eqs.~(\ref{eqconstraintsprojected}) which can be rewritten in the algebraic form,
\begin{subequations}
\label{eqconstraintsdiscrete}
\begin{align}
(q_1^j)^2+ (q_2^j)^2 + (q_3^j)^2 + (q_4^j)^2 - 1 = 0 & \quad \text{for $1 \leq j \leq N$} \label{eqconstraintsdiscretea} \\
\vect{r}^{j+1} - \vect{r}^j - \vect{d}_3^j ds^j = \vect{0} & \quad \text{for $1 \leq j \leq N$,}  \label{eqconstraintsdiscreteb}
\end{align}  
\end{subequations} 
where we used the forward finite difference in Eq.~(\ref{eqderivativeposition}) to approximate $\vect{r}'(s)$. 

The $4N$ geometrical constraints given in Eq.~(\ref{eqconstraintsdiscrete}), together with the $7N+3$ equilibrium equations Eqs.~(\ref{eqequilibriumforcediscrete}) and (\ref{eqequilibriummomentdiscrete}) form the set of algebraic equations describing the constrained equilibrium configuration of the rod represented by the $7N+3$ degrees of freedom $\vect{r}^i$ and $\vect{q}^j$ and the $4N$ Lagrange parameters $\alpha^j$ and $\vect{\mu}^j$. 

We highlight the fact that the only approximations made in the above equations arise from the finite element discretization since the initial continuous formulation is geometrically-exact due to the use of quaternions. Furthermore, it is remarkable to notice that the equilibrium configurations of the extremely twisted and bended elastic rod can be represented by the smooth polynomial equations Eqs.~(\ref{eqequilibriumforcediscrete})-(\ref{eqconstraintsdiscrete}). In the next section, we exploit the particularly smooth nonlinearities of the equilibrium equations by using Asymptotic Numerical Methods (ANM) \cite{damil1990new,cochelin1994path,cochelin1994asymptotic} which are efficient path-following techniques that give access to semi-analytical solution branches of polynomial nonlinear algebraic systems.

\subsection{Asymptotic Numerical Method}\label{ANMmethod}


We now explain and adapt the particular ANM introduced in \cite{cochelin1994path} for solving the equilibrium equations of slender elastic rods described above. This ANM is a perturbation technique allowing for the computation of a large part of a solution branch of quadratic algebraic system of equations with only one stiffness inversion. Applied in a step-by-step manner, one can compute a complex nonlinear branch by a succession of local asymptotic expansions and thus determine a semi-analytical bifurcation diagram. Because of the local analytical representation of the branch within each step, this continuation technique has a number of important advantages when compared to classical predictor-corrector schemes \cite{cochelin1994path}. In particular, the algorithm is fully automatic, remarkably robust, and faster than incremental-iterative methods.

To apply the asymptotic numerical method to the mechanics of elastic rods, we first rewrite the algebraic nonlinear systems of equilibrium equations Eqs.~(\ref{eqequilibriumforcediscrete})-(\ref{eqconstraintsdiscrete}) in the compact form,
\begin{equation}  
\vect{f}\left(\vect{w},\lambda\right)  =  \vect{0}.
\label{eqnonlinearfounctional}
\end{equation}
where $\vect{f}$ is a smooth nonlinear vector valued function in $\mathcal{R}^{11N+3}$ with $N$ the number of elements of the discretized rod, $\lambda$ is a scalar control parameter (usually a mechanical or geometrical parameter of the physical problem such as the rotation or displacement at one end of the rod) and $\vect{w}$ is the vector of unknowns which, in our case reads,
\begin{equation}  
\vect{w} = \tp{\left[\vect{r}^1 \, \vect{q}^1 \ldots \vect{r}^N \, \vect{q}^N \, \vect{r}^{N+1} \, \alpha^1 \, \vect{\mu}^1 \ldots \alpha^N \, \vect{\mu}^N \right]}.
\label{eqstatevector}
\end{equation} 
According to the discretization presented in Section \ref{discretization}, $\vect{w}$ is a vector of size $11N + 3$ which includes the $7N+3$ mechanical degrees of freedom separated into positions $\vect{r}^i$ and quaternions $\vect{q}^j$. Moreover, the $4N$ Lagrange parameters are required to impose the geometrical constraints.

In what follows, in a process that we refer as \emph{recasting}, we now transform Eq.~(\ref{eqnonlinearfounctional}) into a quadratic form, which is a particular framework of the ANM that allows us to formally and systematically write a large class of physical problems including rods \cite{cochelin1994path,cochelin1994asymptotic}. Given the original cubic form of Eq.~(\ref{eqnonlinearfounctional}), this quadratic recast is achieved introducing a new vector of unknowns $\vect{u}$ of size $14N+1$,
\begin{equation}  
\vect{u} = \tp{\left[\vect{w} \, G_1^1 \, G_2^1 \, G_3^1 \ldots G_N^1 \, G_N^2 \, G_N^3 \, \lambda \right]},
\label{eqquadraticstatevector}
\end{equation}
which includes the initial vector of unknowns $\vect{w}$ given in Eq.~(\ref{eqstatevector}), the control parameter $\lambda$ and where we added the $3(N-1)$ flexural and torsional internal torques $G_k^j(\vect{q}^j)$ introduced in Eq.~(\ref{eqinternalmomentcoeffdiscrete}). Using the new vector $\vect{u}$ instead of $\vect{w}$, we can recast the cubic nonlinear vector valued function $\vect{f}\left(\vect{w},\lambda\right)$ given in Eq.~(\ref{eqnonlinearfounctional}) into the quadratic form,
\begin{equation}  
\vect{f}\left(\vect{u}\right)  = \vect{L}_0 + \vect{L}(\vect{u}) + \vect{Q}(\vect{u},\vect{u}) = \vect{0},
\label{eqnonlinearfunctionalquadratic}
\end{equation}
where $\vect{f}\left(\vect{u}\right)$ is a vector in $\mathcal{R}^{14N}$  since we added the $3(N-1)$ nonlinear quadratic equations Eq.~(\ref{eqinternalmomentcoeffdiscrete}) to $\vect{f}\left(\vect{w}\right)$, $\vect{L}_0$ is a constant vector and $\vect{L}({\scriptstyle\bullet})$ and $\vect{Q}({\scriptstyle\bullet},{\scriptstyle\bullet})$ are a linear and bilinear vector valued operators, respectively. The expression of $\vect{f}\left(\vect{u}\right)$, representing the equilibrium equations of our inextensible and unshearable elastic rod given in Eqs.~(\ref{eqequilibriumforcediscrete})-(\ref{eqconstraintsdiscrete}), is provided in \ref{quadfu}. In Section \ref{sec5} where we will apply our method to the quasi-static writhing of a double-clamped elastic rod, we will illustrate, by way of example, how to introduce the boundary conditions and the control parameter in the vector $\vect{f}\left(\vect{u}\right)$ of Eq.~(\ref{eqnonlinearfunctionalquadratic}).

We can now proceed and compute the solutions $\vect{u}$ of the set of quadratic equations Eqs.~(\ref{eqnonlinearfunctionalquadratic}) with the asymptotic numerical method. This technique is based on the perturbation of the vector of unknowns $\vect{u}$ in terms of a path-parameter $a$ in the form of the asymptotic expansion,
\begin{equation}  
\vect{u}(a) = \vect{u}_0 + \sum_{p=1}^{m}a^p\vect{u}_p,
\label{eqasymptoticdev}
\end{equation}
where $\vect{u}_0$ is the starting fixed point, solution of Eq.~(\ref{eqnonlinearfunctionalquadratic}), $m$ is the truncation order of the power series and $a$ is the path-parameter which will be formally defined below. Replacing $\vect{u}(a)$ by its asymptotic expansion Eq.~(\ref{eqasymptoticdev}) in the quadratic form Eq.~(\ref{eqnonlinearfunctionalquadratic}), we obtain the quadratic Taylor series in the neighborhood of $\vect{u}_0$,
\begin{align}
\vect{f}\left(\vect{u}(a)\right) & = \vect{L}_0 + \vect{L}(\vect{u_0}) + \vect{Q}(\vect{u_0},\vect{u_0}) \nonumber \\
                                        & + a \left[\vect{L}(\vect{u}_1) + 2\vect{Q}(\vect{u}_0,\vect{u}_1)\right] \label{eqdevlimitfunctional}\\
                                        & + \sum\nolimits_{p=1}^{m} a^p\left[\vect{L}(\vect{u}_p) + 2\vect{Q}(\vect{u}_0,\vect{u}_p) + \sum\nolimits_{i=1}^{p-1}\vect{Q}(\vect{u}_i,\vect{u}_{p-i}) \right]. \nonumber
\end{align}  

Recalling that $\vect{u}_0$ is a solution of Eq.~(\ref{eqnonlinearfunctionalquadratic}), we can rewrite Eq.~(\ref{eqdevlimitfunctional}) in the form of a power series of a quadratic vector  valued function $\vect{f}_p$ of size $14N$,
\begin{equation}  
\vect{f}\left(\vect{u}(a)\right)  = a\vect{f}_1 + a^2\vect{f}_2 + \ldots + a^m\vect{f}_m = \vect{0}.
\label{eqdevlimitfunctionalcompact}
\end{equation}
Since Eq.~(\ref{eqdevlimitfunctionalcompact}) has to be verified for every value of $a$, we need $\vect{f}_p = \vect{0}$ for every order $p \leq m$.  This leads to $m$ linear systems in $\vect{u}_p$ in the form,
\begin{subequations}
\label{eqlinearsystems}
\begin{equation}  
\forall p \in [1 \ldots m], \quad \frac{\partial \vect{f}}{\partial \vect{u}}\bigg|_{\vect{u}_0} \vect{u}_p = \vect{f}_p^{nl},
\label{eqlinearsystemsa}
\end{equation}
where, due to the particular quadratic form of Eq.~(\ref{eqnonlinearfunctionalquadratic}), the Jacobian matrix of $\vect{f}(\vect{u})$ evaluated at the initial solution vector $\vect{u}_0$ reads,
\begin{equation}
\frac{\partial \vect{f}}{\partial \vect{u}}\bigg|_{\vect{u}_0} \vect{u}_p = \vect{L}(\vect{u}_p) + 2\vect{Q}(\vect{u}_0,\vect{u}_p), \label{eqlinearsystemsb}
\end{equation}
and the nonlinear vector $\vect{f}_p^{nl}$ on the right-hand side of Eq.~(\ref{eqlinearsystemsa}) consists of a quadratic sum that only depend on the previous order,
\begin{align}
\vect{f}_p^{nl} & = 0 & \text{for $p = 1$} \label{eqlinearsystemsc} \\
\vect{f}_p^{nl} & = - \sum\nolimits_{i=1}^{p-1}\vect{Q}(\vect{u}_i,\vect{u}_{p-i}) & \text{for $1 < p \leq m$.} \label{eqlinearsystemsc}
\end{align}  
\end{subequations} 
The original nonlinear problem in Eq.~(\ref{eqnonlinearfunctionalquadratic}) has thereby been reduced to a definite set of $m$ linear systems given in Eq.~(\ref{eqlinearsystemsa}) where the matrix on the left-hand side is identical for each order. 

\begin{figure}[t!]
\includegraphics[width=\columnwidth]{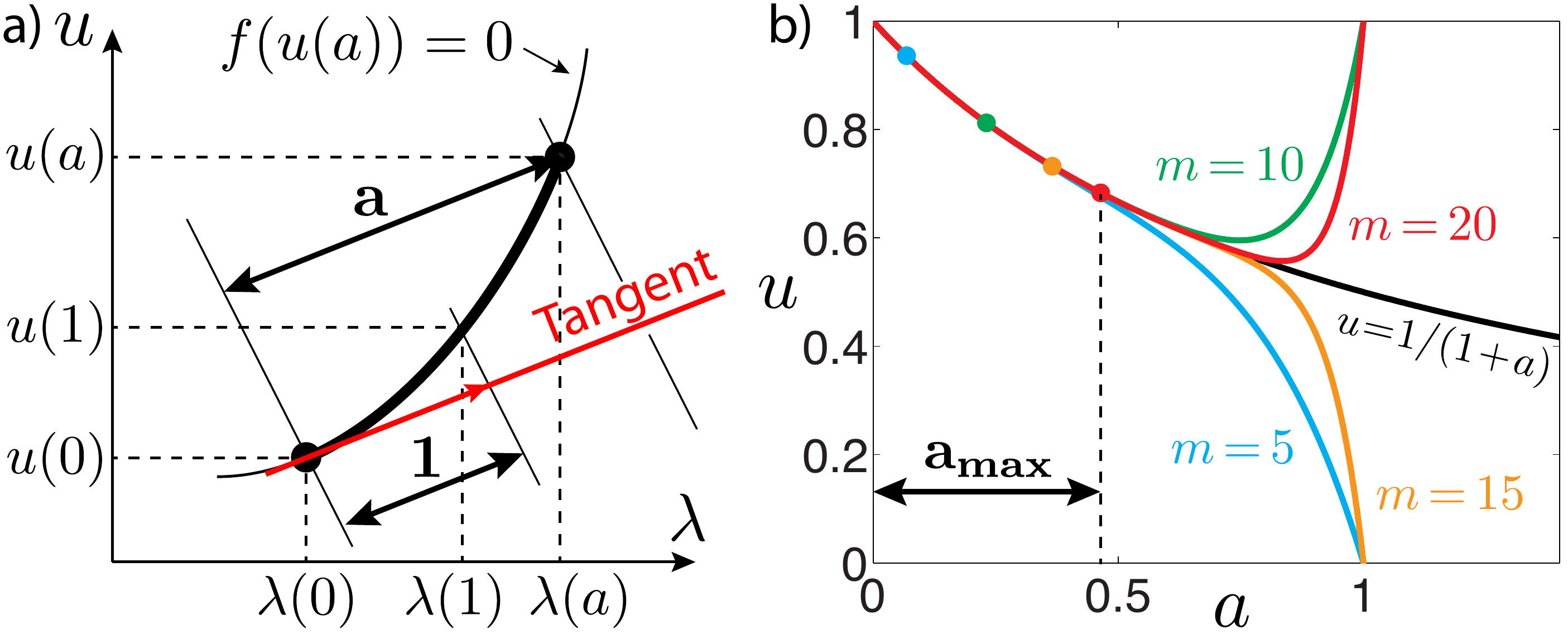}
\caption{\label{fig:ANM_method} (a) The path parameter $a$ is identified as the projection of the solution branch on the tangent vector. This projection is normalized by the length of the tangent vector $\left[u(1) \, \lambda(1)\right]$ which is set to unity. (b) Behavior of a power series close to the radius of convergence. The considered nonlinear unidimensional equation is $f\left(u(a)\right) = u(1+a) - 1 = 0$. Its unique solution $u(a) = 1/(1+a)$, represented as solid line, can be expanded asymptotically as $u(a) \approx 1 - a + a^2 + \ldots + (-1)^m a^m + (-1)^{m+1} a^{m+1}$. The asymptotic expansions for $m = 5$, $10$, $15$ and $20$ are represented as different lines. We see that $u(a)$ can be approximated up to the radius of convergence $a = 1$. Applying our ANM method, the step length calculated through Eqs.~(\ref{eqarclengthmaxidemoa})-(\ref{eqlengthstep}) would give $a_{max} =  \varepsilon^{1/(m+1)} = 0.4642$ with $\varepsilon = 1\times10^{-7}$ and $m = 20$.}
\end{figure}

However, each linear system in Eqs.~(\ref{eqlinearsystems}) is, so far, under-determined since the dimension of $\vect{f}$ is $14N$ whereas the size of the vector of state variable at order $p$, $\vect{u}_p$, is $14N+1$. The remaining equation is provided by the definition of the path parameter $a$ as defined in \cite{cochelin1994path}. 
We consider a measure that includes the entire set of physical unknowns and that is also robust towards limit and bifurcation points, i.e. an arc-length measure. Mathematically, we identify the path parameter $a$ as the projection of the vector of state variables increment $\vect{u} - \vect{u}_0$ on the normalized tangent vector $\vect{u}_1$ [see Fig.~\ref{fig:ANM_method}(a)],
\begin{equation}  
a = \tp{\left(\vect{u}(a) - \vect{u}_0\right)}\vect{u}_1.
\label{eqpathparameterdefinition}
\end{equation}
Replacing $\vect{u}(a)$ by its asymptotic expansion, Eq.~(\ref{eqasymptoticdev}), in Eq.~(\ref{eqpathparameterdefinition}), we obtain,
\begin{equation}  
a\tp{\vect{u}_1}\vect{u}_1 - a + a^2\tp{\vect{u}_1}\vect{u}_2 + \ldots + a^m \tp{\vect{u}_1}\vect{u}_m = 0.
\label{eqdevpathparameter}
\end{equation}
Verifying Eq.~(\ref{eqdevpathparameter}) at every power of $a$ provides us with the supplementary equations at every $p$,
\begin{equation}
\label{eqpathparametersuppequations}
\tp{\vect{u}_1}\vect{u}_p = \delta_{p1} \quad \text{for $1 \leq p \leq m$,}
\end{equation} 
where $\delta_{p1}$ is the Kronecker delta, $\delta_{11} = 1$ at the first order $p = 1$ and it is zero otherwise. 

Finally, the original nonlinear problem in Eq.~(\ref{eqnonlinearfunctionalquadratic}) has now been transformed in the well-posed $m$ linear systems in $\mathcal{R}^{14N+1}$,
\begin{equation}  
\left[\begin{array}[c]{c}
\vspace{0.2cm}
\frac{\partial \vect{f}}{\partial \vect{u}}\bigg|_{\vect{u}_0} \\
\tp{\vect{u}_1}
\end{array}
\right]\vect{u}_p =
\left\{\begin{array}[c]{c}
\vspace{0.6cm}
\vect{f}_p^{nl}  \\
\delta_{p1}
\end{array}
\right\} \quad \text{for $1 \leq p \leq m$,}
\label{eqwellposedlinearsystems}
\end{equation}
with a unique solution $\vect{u}_p$ that we can solve iteratively since each of these vectors is defined with the solution of the previous order according to the definition of $\vect{f}_p^{nl}$ given in Eq.~(\ref{eqlinearsystems}). In this linearized numerical problem, the only matrix to inverse is the one on the left-hand side of Eq.~(\ref{eqwellposedlinearsystems}) since it is the same at every order $p$. This is in striking contrast with classical predictor-corrector methods where one needs to actualize the Jacobian for every linear systems \cite{crisfield1991nonlinear}.  

For practical purposes, one will inverse the matrix and compute the unknown $\vect{u}_1$ at first order independently and then compute the higher orders $\vect{u}_p$ for $1 < p \leq m$ from the well-defined systems in Eq.~(\ref{eqwellposedlinearsystems}). 

Once each $\vect{u}_p$ has been found, we still have to estimate the validity domain of the asymptotic expansion since Eq.~(\ref{eqnonlinearfunctionalquadratic}) can only be true for values of the perturbation parameters $a$ inside the radius of convergence of the power series given in Eq.~(\ref{eqasymptoticdev}) [see Fig.~\ref{fig:ANM_method}(b)]. A simple, robust and accurate way of calculating an approximation of the convergence radius $a_{max}$, explained in detail in \cite{cochelin1994path}, is to assume that a solution branch is acceptable as long as the norm of the nonlinear $\left(14N+1\right)$-dimensional vector field $\vect{f}\left(\vect{u}(a)\right)$ is less than a tolerance criterion $\varepsilon$,
\begin{equation}
\label{eqresiduesmall}
\forall a \in \left[0 \, a_{max}\right], \quad \|\vect{f}\left(a\right)\| < \varepsilon,
\end{equation}
where $\varepsilon$ determines the accuracy of our numerical results. We have computed the $\vect{u}_p$ according to the power series expansion of $\vect{f}\left(a\right)$ given in Eq.~(\ref{eqdevlimitfunctionalcompact}) so that the norm of $\vect{f}$ is zero up to the truncation order $m$. Consequently, the residue of this series is given by the norm of $\vect{f}\left(a\right)$ for $p > m$. Assuming that the order $m+1$ dominates in the residue, we obtain the relation between the norm of $\vect{f}\left(a\right)$ and the vector at the order $m+1$,
\begin{equation}  
\|\vect{f}\left(a\right)\| \approx a^{m+1}\|\vect{f}_{m+1}\|
\label{eqarclengthmaxidemoa}
\end{equation}
where $\vect{f}_{m+1} =  \frac{\partial \vect{f}}{\partial \vect{u}}\big|_{\vect{u}_0} \vect{u}_{m+1} - \vect{f}_{m+1}^{nl}$ according to Eqs.(\ref{eqdevlimitfunctionalcompact})-(\ref{eqlinearsystems}). Replacing $\|\vect{f}\left(a\right)\|$ by its definition Eq.~(\ref{eqarclengthmaxidemoa}) in Eq.~(\ref{eqresiduesmall}) leads to,
\begin{equation}
\|\vect{f}\left(a\right)\| < \varepsilon \quad \Leftrightarrow \quad a < \left(\frac{\varepsilon}{\|\vect{f}_{m+1}\|}\right)^{\frac{1}{m+1}}.
\label{eqarclengthmaxidemob}
\end{equation}
which sets an upper limit to the path-parameter $a$. While truncating the asymptotic  series $\vect{u}(a)$ given in Eq.~(\ref{eqasymptoticdev}) at the order $m$, we implicitly assumed that $\vect{u}_{m+1} = \vect{0}$ so that $\vect{f}_{m+1} = \vect{f}_{m+1}^{nl}$. Replacing $\vect{f}_{m+1}$ by the nonlinear term $\vect{f}_{m+1}^{nl}$ in Eq.~(\ref{eqarclengthmaxidemob}), we obtain an estimation of the maximum step length,
\begin{equation}
a_{max} = \left(\frac{\varepsilon}{\|\vect{f}_{m+1}^{nl}\|}\right)^{\frac{1}{m+1}}.
\label{eqlengthstep}
\end{equation}
For practical purposes, when applying Eq.~(\ref{eqlengthstep}) to any quadratic vector valued function $\vect{f}(\vect{u}(a))$, we found that $\vect{f}\left(\vect{u}(a_{max})\right) \approx \varepsilon$. In general, the power series in Eq.~(\ref{eqasymptoticdev}) converges slowly, close to the radius of convergence \cite{cochelin1994asymptotic}. Decreasing $\varepsilon$ leads to a diminishing of $a_{max}$ but, more importantly, results in an increased accuracy of the computed series. An optimal value of $m$ and $\varepsilon$ in terms of convergence of the asymptotic series, accuracy of the solution given by Eq.~(\ref{eqresiduesmall}) and size of $a_{max}$ is found empirically for the following range of parameters \cite{cochelin1994asymptotic}: $\varepsilon = 1\times10^{-7}$ and $15 \leq m \leq 20$. 

\begin{figure}[t!]
\includegraphics[width=\columnwidth]{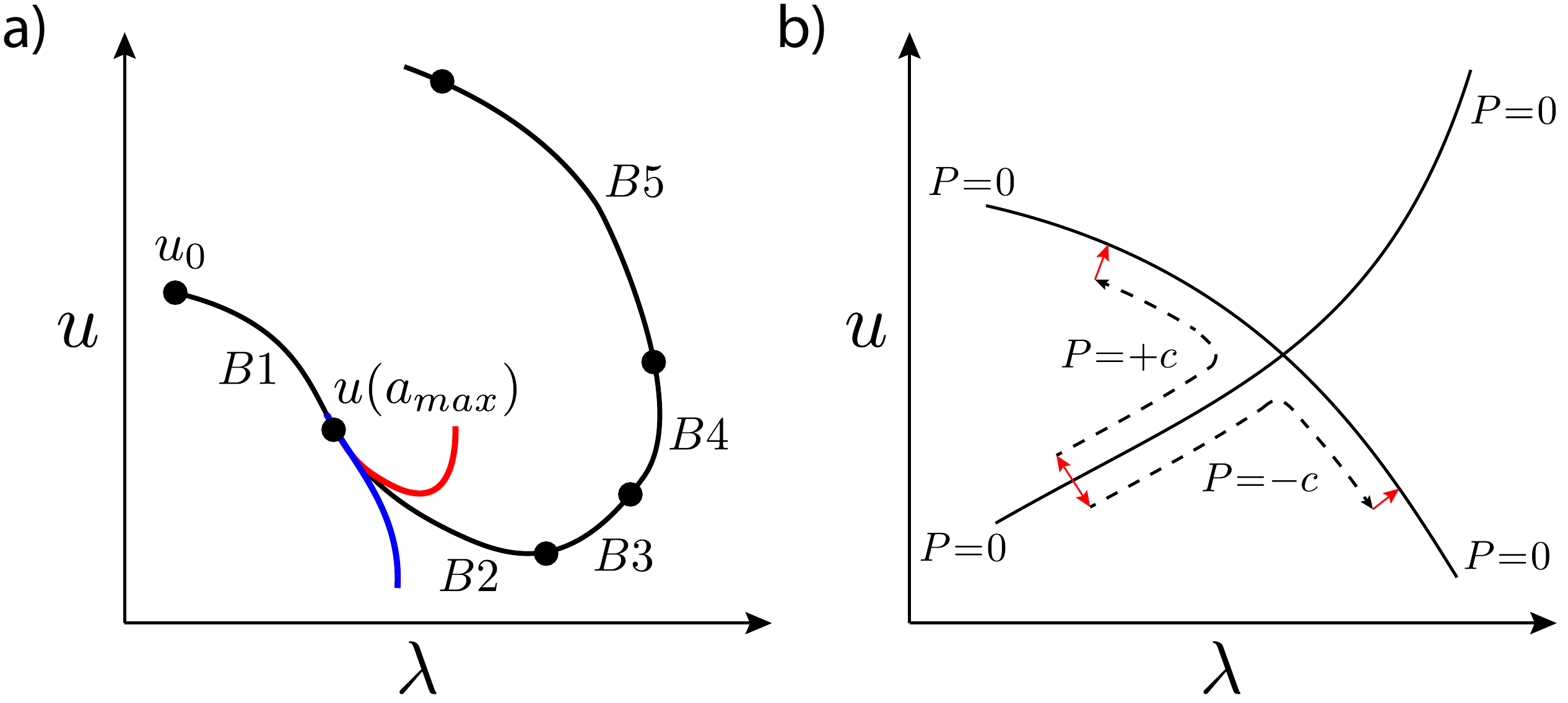}
\caption{\label{fig:ANM_continuation} (a) Schematic five continuation steps with the asymptotic numerical method. The step length of each portion of branch is determined \emph{a posteriori} by analyzing the validity of the asymptotic solution. The highly nonlinear solution is known analytically for each portion. (b) Branch switching through the perturbation method given in Eq.~(\ref{eqnonlinearfunctionalquadraticperturbed}). Changing the sign of the intensity of perturbation, $c$, allows us to explore all the branches after a bifurcation point.}
\end{figure}

The power series expansion given in Eq.~(\ref{eqasymptoticdev}) and computed with the $m$ linear systems in Eq.~(\ref{eqwellposedlinearsystems}), together with the maximum step size $a_{max}$ given by Eq.~(\ref{eqlengthstep}) define a portion of the nonlinear equilibrium branches of the slender elastic rod in terms of a given control parameter $\lambda$. The next step of our calculation, the continuation of the solution branch, is now computed by applying the present asymptotic numerical method taking $\vect{u}(a_{max})$ as the new starting equilibrium $\vect{u}_0$ of the new portion. A complete solution branch is therefore constructed as a succession of semi-analytical portions in the form of Eq.~(\ref{eqasymptoticdev}), whose length is automatically determined through the estimation of the convergence radius of each power series as sketched in Fig.~\ref{fig:ANM_continuation}(a). Unlike classical predictor-corrector methods \cite{riks1979incremental,doedel1981auto}, our step length is adaptive; it is naturally large for weakly nonlinear solutions and becomes shorter when strong nonlinearities occur. As a consequence, the automatization of the continuation method is significantly easier and more robust than with standard predictor-corrector methods.

One would expect that the residue $\|\vect{f}\left(u(a)\right)\|$ would increase progressively at every continuation step so that the accuracy of the new starting equilibrium $\vect{u}_0$ would gradually decrease \cite{cochelin1994asymptotic}. In practice however, it is rare to see the residue increase up to $10\varepsilon$, especially given the smooth nature of the nonlinearities of our equilibrium equations Eqs.~(\ref{eqequilibriumforcediscrete})-(\ref{eqconstraintsdiscrete}) for thin elastic rods. In Section \ref{sec5}, where we implement this continuation method to a series of specific test-case problems, all the bifurcation diagrams are computed with a residue smaller than $\varepsilon = 1\times10^{-7}$, with no correction step (note that a correction step may be necessary in the general ANM framework in the case of non-polynomial nonlinearities \cite{Karkar2013968}).    

When the accuracy $\varepsilon$ is set to be small enough by the user, the ANM is able to follow the branch whenever bifurcation are encountered \cite{baguet2003behaviour}. This is a remarkably robust property for a path-following algorithm, especially when compared to predictor-corrector techniques which typically would systematically bifurcate because of the discrete nature of their continuation steps. Nevertheless, we now need a special procedure to switch branches in order to determine the full bifurcation diagram. A classic strategy is to slightly modify the original equilibrium equations Eq.~(\ref{eqnonlinearfunctionalquadratic}) by adding a low-norm perturbation vector,
\begin{equation}  
\vect{f}_P\left(\vect{u}\right)  = \vect{L}_0 + \vect{L}(\vect{u}) + \vect{Q}(\vect{u},\vect{u}) + c\vect{P},
\label{eqnonlinearfunctionalquadraticperturbed}
\end{equation}
where $\vect{f}_P\left(\vect{u}\right)$ is the perturbed problem, $\vect{P}$ a normalized vector of constant random numbers and $c$ is the intensity of the perturbation. This additional perturbation procedure transforms the exact bifurcation into a perturbed bifurcation [see Fig.~\ref{fig:ANM_continuation}(b)]. The idea is to use the perturbed branch to bifurcate on the non-crossing branch \cite{allgower2003introduction}. Changing the sign of $c$ allows us to explore a symmetrical quasi-bifurcation as represented in Fig.~\ref{fig:ANM_continuation}(b) with the perturbed branches $P = +c$ and $P = -c$. Finally, in order to transition from the original to the perturbed problem, or vice versa, a correction step is mandatory (any predictor-corrector methods would be efficient since the perturbed solutions are very close to exact ones). The combination of different (positive and negative) values of the intensity of the perturbation and several correction steps allows one to explore the full bifurcation diagram of the slender elastic rod described by the equilibrium equations Eqs.~(\ref{eqequilibriumforcediscrete})-(\ref{eqconstraintsdiscrete}).

So far, we have presented the ANM method in the context of thin elastic rods. We can now compute the bifurcation diagrams of our slender elastic rod under various mechanical and geometrical environments. However, the final crucial step of determining the stability of the solution branches is still missing, which is the focus of the following section.

\subsection{Stability analysis}\label{secstab}

Determining the local stability of equilibrium branches is crucial for the physical understanding of the mechanical behavior of slender elastic rods, one of the main motivations being that locally unstable branches cannot be observed experimentally, and must therefore be classified. Another advantage for gaining knowledge on the stability of a solution is that the loss of local stability is often associated with a bifurcation point. Assessing the stability is then useful to detect bifurcation points and navigate through the bifurcation diagram as illustrated in Fig.~\ref{fig:ANM_continuation}(b). 

The equilibrium equations on their own are not sufficient to determine the local stability of the solutions; we also need to compute wether the solution is a local minimum or maximum of the system's energy. For practical purposes, we need to derive the second-order conditions of the geometrically constrained energy; a theoretical and numerical procedure that is well established \cite{luenberger1973introduction,luenberger2008linear}. 


In the previous sections, we have shown how to compute and follow the branches of solutions of the nonlinear algebraic equilibrium equations given in Eqs.~(\ref{eqequilibriumforcediscrete})-(\ref{eqconstraintsdiscrete}). Since the resulting computed bifurcation diagrams are semi-analytical, we are therefore able to evaluate the solution $\vect{u}_e$ given in Eq.~(\ref{eqquadraticstatevector}) at any finite value of the control parameter $\lambda_e$. Let $\vect{x}_e$, $\vect{\alpha}_e$ and $\vect{\mu}_e$ be the vectors of degrees of freedom and Lagrange multipliers respectively, associated with the vector solution $\vect{u}_e$, such that,
\begin{subequations}
\label{eqddlsolution}
 \begin{align} 
\vect{x_e} & = \tp{\left[r_x^1 \, r_y^1 \, r_z^1 \, q_1^1 \, q_2^1 \, q_3^1 \, q_4^1 \ldots r_x^N \, r_y^N \, r_z^N \, q_1^N \, q_2^N \, q_3^N \, q_4^N \, r_x^{N+1} \, r_y^{N+1} \, r_z^{N+1} \right]},  \label{eqddlsolutiona} \\
\vect{\alpha}_e & = \tp{\left[\alpha^1 \, \alpha^2 \ldots \alpha^N \right]} , \label{eqddlsolutionb} \\
\vect{\mu}_e & = \tp{\left[\mu_x^1 \, \mu_y^1 \, \mu_z^1 \ldots \mu_x^N \, \mu_y^N \, \mu_z^N\right]}.  \label{eqddlsolutionc}
\end{align} 
\end{subequations}
The vector $\vect{x_e}$ is a solution of the equilibrium equations Eqs.~(\ref{eqequilibriumforcediscrete})-(\ref{eqequilibriummomentdiscrete}) for $\lambda = \lambda_e$ that satisfies the functional geometrical constraints given in Eq.~(\ref{eqconstraintsdiscrete}). Rewriting Eqs.~(\ref{eqequilibriumforcediscrete})-(\ref{eqconstraintsdiscrete}) in an energy minimization framework, $\vect{x}_e$ is the actual solution of the $n = 7N+3$-dimensional constraint minimization problem,
\begin{subequations}  
\label{eqminimizationproblem}
\begin{equation} 
\vect{\nabla}\left(\mathcal{E}_{e}\left(\vect{x}_e\right) + \mathcal{W}\left(\vect{x}_e\right)\right) + \sum_{i=1}^N\vect{\alpha}_e^i\vect{\nabla}\vect{\mathcal{C}}_{\alpha}^i + \sum_{j=1}^{3N}\vect{\mu}_e^j\vect{\nabla}\vect{\mathcal{C}}_{\mu}^j = \vect{0},
\label{eqminimizationproblema}
\end{equation}
subject to the $m = 4N$ functional constraints,
\begin{align}
\mathcal{C}_{\alpha}^j\left(\vect{x_e}\right) & = \vect{q}^j\vect{q}^j - 1 = 0, \label{eqminimizationproblemb}\\
\mathcal{C}_{\mu x}^j\left(\vect{x_e}\right) & = r_x^{j+1} - r_x^j - 2q_1^jq_3^j - 2q_2^jq_4^j  = 0, \label{eqminimizationproblemc}\\
\mathcal{C}_{\mu y}^j\left(\vect{x_e}\right) & = r_y^{j+1} - r_y^j - 2q_2^jq_3^j + 2q_1^jq_4^j = 0, \label{eqminimizationproblemd}\\
\mathcal{C}_{\mu z}^j\left(\vect{x_e}\right) & = r_z^{j+1} - r_z^j + (q_1^j)^2 + (q_2^j)^2 - (q_3^j)^2 - (q_4^j)^2 = 0, \label{eqminimizationprobleme}
\end{align}  
\end{subequations}
for the positional nodes $1 \leq j \leq N$. In Eq.~(\ref{eqminimizationproblem}), $\vect{\nabla}$ is the gradient operator \footnote{For a real-valued function $f \in C^1$ on $\mathcal{R}^n$ such that $f(\vect{x}) = f(\vect{x}_1, \vect{x}_2, \ldots, \vect{x}_n)$, we define the gradient of $f$ to be the $n$-dimensional vector,
\begin{equation*}
\vect{\nabla} f\left(\vect{x}\right) = \tp{\left[\frac{\partial f(\vect{x})}{\partial\vect{x}_1} \, \frac{\partial f(\vect{x})}{\partial\vect{x}_2} \ldots \frac{\partial f(\vect{x})}{\partial\vect{x}_n}\right]}.
\end{equation*}
}, the real-valued function $\mathcal{E}_{e}\left(\vect{x}_e\right)$ measures the amount of elastic energy stored in the rod at equilibrium,
\begin{equation}  
\mathcal{E}_{e}\left(\vect{x_e}\right) = \sum_{k=1}^3\frac{E_kI_k}{2}\sum_{j=1}^{N-1}\left(\mat{B}_k\left(\vect{q}^j+\vect{q}^{j+1}\right)\frac{1}{ds}\left(\vect{q}^{j+1}-\vect{q}^j\right) - \hat{\kappa}_k^j\right)^2ds^j,
\label{eqbendingenergyquaternionsbis}
\end{equation}
and $\mathcal{W}\left(\vect{x}_e\right)$ quantifies the total work of external forces and moments,
\begin{multline}  
\mathcal{W}\left(\vect{x_e}\right) = \vect{P}^0\vect{r}^1 + \vect{M}^0\vect{q}^1 + \vect{P}^L\vect{r}^{N+1} + \vect{M^L}\vect{q}^N + \\
                                   \sum_{i=2}^{N} \vect{p}^i \vect{r}^ids + \sum_{j=2}^{N-1}\vect{m}^j \vect{q}^j ds.
\label{eqexternalworkdiscretebis}
\end{multline}
Finally, the vectors $\vect{\mathcal{C}}_{\alpha} = \tp{\left[\mathcal{C}_{\alpha}^1 \, \mathcal{C}_{\alpha}^2 \ldots \mathcal{C}_{\alpha}^N\right]}$ and $\vect{\mathcal{C}}_{\mu} = \tp{\left[\mathcal{C}_{\mu x}^1 \, \mathcal{C}_{\mu y}^1 \, \mathcal{C}_{\mu z}^1 \ldots \mathcal{C}_{\mu x}^N \, \mathcal{C}_{\mu y}^N \, \mathcal{C}_{\mu z}^N\right]}$ in Eq.~(\ref{eqminimizationproblema}) represent the geometrical constraints that ensure the norm of quaternions to be one and the inextensibility of the oriented Cosserat rod respectively, and should be very close to zero at equilibrium. 

According to the necessary and sufficient first order conditions of constrained minimization problems \cite{luenberger1973introduction,luenberger2008linear}, the vector solution $\vect{x}_e$ is a local extremum (a minimum or maximum) of the total energy $\mathcal{E}(\vect{x_e}) = \mathcal{E}_e(\vect{x_e}) + \mathcal{W}(\vect{x_e})$ subject to the $m$ constraints in Eqs.~(\ref{eqminimizationproblemb})-(\ref{eqminimizationprobleme}). Supposing also that the $n \times n$ matrix,
\begin{equation}  
\mat{L}(\vect{x}_e) = \vect{\nabla}^2\left(\mathcal{E}_{e}\left(\vect{x}_e\right) + \mathcal{W}\left(\vect{x}_e\right)\right) + \sum_{i=1}^N\vect{\alpha}^i\vect{\nabla}^2\vect{\mathcal{C}}_{\alpha}^i + \sum_{j=1}^{3N}\vect{\mu}^j\vect{\nabla}^2\vect{\mathcal{C}}_{\mu}^j,
\label{eqsecondordercondition}
\end{equation}
where $\vect{\nabla}^2$ is the Hessian operator \footnote{We define the Hessian of $f$ at $\vect{x}$ ($f \in C^2$) to be the $n \times n$ symmetric matrix denoted $\vect{\nabla}^2 f\left(\vect{x}\right)$,
\begin{equation*}
\vect{\nabla}^2 f\left(\vect{x}\right) = \left[\frac{\partial^2 f(\vect{x})}{\partial\vect{x}_i \partial\vect{x}_j}\right].
\end{equation*}}
, is positive definite on the $m$-dimensional subspace $M = \left\{\vect{y} : \vect{\nabla}\vect{h}(\vect{x}_e)\vect{y} = \vect{0}\right\}$ with $\vect{h}(\vect{x}_e) = \tp{\left[\vect{\mathcal{C}}_{\alpha} \, \vect{\mathcal{C}}_{\mu}\right]} $, that is, for $\vect{y} \in M$ and $\vect{y} \neq \vect{0}$ that holds $\tp{\vect{y}}\mat{L}(\vect{x}_e)\vect{y} > 0$, then, according to the second-order necessary and sufficient conditions, $\vect{x}_e$ is a strict local minimum of $\mathcal{E}(\vect{x_e})$ subject to $\vect{\mathcal{C}}_{\alpha} = \vect{0}$ and $\vect{\mathcal{C}}_{\mu}  = \vect{0}$.

The matrix $\mat{L}(\vect{x}_e)$ is the matrix of second partial derivatives, with respect to $\vect{x}$, of the discrete counterpart of the Lagrangian given in Eq.~(\ref{eqvariationprinciple}). When restricted to the subspace $M$ that is tangent to the constraint surface and which we denote by $\mat{L}_M$, $\mat{L}(\vect{x}_e)$ plays the role in second-order conditions directly analogous to that of the Hessian of the objective function in the unconstrained case \cite{luenberger1973introduction,luenberger2008linear}. The eigenvalues, $\sigma_i$, and associated eigenvectors, $\vect{y}_i$, of $\mat{L}_M$, determine the local stability of the solutions of the constrained minimization problem. Mathematically, $\mat{L}_M$ is a $\left(n-m\right) \times \left(n-m\right)$ matrix defined, at each equilibrium point $\vect{x}_e$, as,
\begin{equation}  
\mat{L}_M(\vect{x}_e) = \tp{\left(\ker(\vect{\nabla}\vect{h})\right)} \mat{L} \left(\ker(\vect{\nabla}\vect{h})\right),
\label{eqLM}
\end{equation}
where $\ker({\scriptstyle\bullet})$ denotes the kernel operator.
Analyzing the $n-m$ eigenvalues $\sigma_i$ gives us  information on the behavior of the associated perturbation $\vect{y}_i(t)$ in the neighborhood of the equilibrium $\vect{x}_e$. According to Lyapunov's theorem \cite{bavzant2010stability,guckenheimer1983nonlinear}:  
\begin{itemize}
\item If $\sigma_i > 0$ for all $i \in \left[1 \ldots n-m\right]$, all the perturbations vanish, $\vect{y}(t) \rightarrow \vect{0}$, when $t \rightarrow +\infty$, and the equilibrium is \emph{locally asymptotically stable}.

\item If one index $i \in \left[1 \ldots n-m\right]$ exists, for which $\sigma_i < 0$, one perturbation diverges, $\vect{y}(t) \rightarrow +\infty$ when $t \rightarrow +\infty$, and the equilibrium is \emph{locally unstable}.

\item If $\sigma_i \geq 0$ for all $i \in \left[1 \ldots n-m\right]$ and if there exists one index $k$ such that $\sigma_k = 0$, the first order is insufficient to draw conclusions on the local stability of the equilibrium. In that case, a perturbation at higher order is necessary. 
\end{itemize}

Applying this method to a sufficient number of fixed points $\vect{x}_e$ along the equilibrium branches, computed with the previous ANM method, allows us to determine the stability of the bifurcation diagram. The previous finite element discretization presented in Section \ref{discretization}, together with the ANM algorithm described in Section \ref{ANMmethod} and the previous stability method, completes the semi-analytical continuation technique that we developed to compute and follow the equilibrium branches and the stability of an inextensible slender elastic rod undergoing extreme displacements and rotations. The combination of the conciseness and relative simplicity of our method offer the opportunity for it to be implemented in any programming language. In the following, we briefly present MANlab \cite{arquier2007methode,manlab}, an open-source bifurcation analysis software that provides a convenient framework to implement the previous numerical methods.

\subsection{MANlab: an open-source bifurcation analysis software}

MANlab is an interactive software package for the continuation and bifurcation analysis of algebraic systems, based on ANM continuation, and first released in 2009 \cite{arquier2007methode}. Thanks to the implementation of most of the ANM equations in MATLAB using an object-oriented approach \cite{manlab}, MANlab makes it simpler for the user to solve the system of Eqs.~(\ref{eqequilibriumforcediscrete})-(\ref{eqconstraintsdiscrete}) and the stability of the solutions given by the second-order condition through Eq.~(\ref{eqsecondordercondition}). MANlab has a graphical user interface (GUI) with buttons, on-line inputs and graphical windows for generating, displaying and analyzing the bifurcation diagram and the solutions of the system. A unique identifying feature, when compared with other continuation codes, is that its computational efficiency, highlighted above, allows for interactive control of the continuation process. The full interactive and semi-automatic procedure consists of computation of a portion of a branch, choice of a new branch at a bifurcation point, reverse direction of continuation on the same branch, jump capability between solutions, visualization of user-defined quantities at a particular solution point, selection and deletion of a branch, or of one of its portion, possibility of correction step with a Newton-Raphson method and determination of the local stability of the solution.

To enter the system of equations, the user simply has to provide the three vector valued Matlab functions corresponding to the constant, linear and quadratic operators $\vect{L}_0$, $\vect{L}(\vect{u})$, and $\vect{Q}(\vect{u},\vect{u})$ given in Eq.~(\ref{eqnonlinearfunctionalquadratic}). To assess the local stability at each computed solution point in MANlab \cite{lazarus2010harmonica}, one can also provide the Hessian of the constrained Lagrangian restricted to $M$, $\mat{L}_M(\vect{x}_e)$  given in Eq.~(\ref{eqLM}) and the package will automatically compute the eigenvalues of the linearized problem according to the previous section. Thanks to the flexibility offered by the MATLAB environment, users become rapidly familiar with MANlab. Calling of external routines such as finite elements codes is also possible. 

In the following section, we validate our semi-analytical continuation method by using the MANlab package to simulate a precision model experiment; the quasi-static writhing of a double-clamped slender elastic rod, which equilibria are solutions of the discrete equilibrium equations given in Eqs.~(\ref{eqequilibriumforcediscrete})-(\ref{eqconstraintsdiscrete}). 

\section{Following the equilibria of an extremely twisted elastic rod}\label{sec5}

Having introduced the general theoretical and numerical framework to continue the equilibria and stability of slender elastic rods, we proceed by implementing the specific problem of the writhing (extreme twisting) of a clamped elastic rod. Even though this fundamental problem appears
seemingly simple, it can display an array of complex behavior with intricate bifurcation diagrams and has received significant attention in the literature~\cite{thompson1996helix,goriely1997nonlinear1,van2000helical,goyal2008non,goriely1997nonlinear2,goriely1997nonlinear3,goriely1998nonlinear4}. The writhing of an elastic rod is therefore an ideal scenario to challenge our theoretical and computational framework by contrasting the numerical results with our own precision model experiments that were especially developed for the testing and validation of our continuation method.

In this Section, we first present our apparatus and model experiments which consist of quasi-statically increasing the rotation angle at one end of a slender elastic rod fixed between two concentrically aligned horizontal clamps. One of the originalities of our experiments is that we fabricate our own elastic rods, enabling us to accurately target their material and geometrical properties. In particular, we have full control in setting their intrinsic natural curvatures. After describing how to account for the kinematic boundary conditions and control parameter specific to this writhing problem in the numerical model described in Section \ref{sec4}, we compare a series of  experimental and numerical results for two different elastic rods: a straight rod with no natural curvature and a curved rod.

\subsection{Manufacturing of rods and experimental apparatus}


\begin{figure}[b!]
\includegraphics[width=\textwidth]{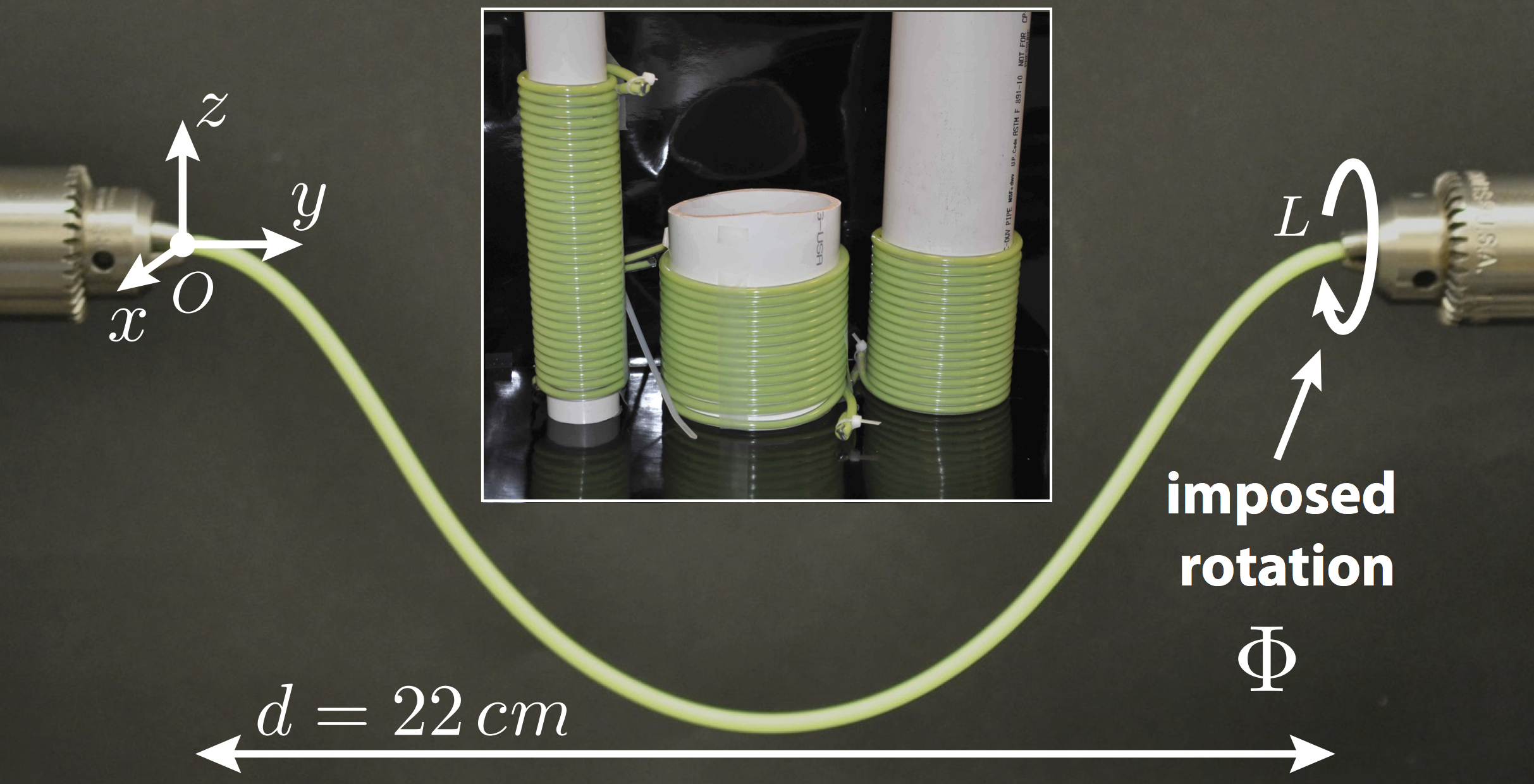}
\caption{\label{fig:experimental_setup}  The writhing experiment. A $L = 30$ cm long elastic rod with a circular cross section of radius $1.55$ mm, Young's modulus $E = 1000$ KPa, volumic mass $\rho = 1200$ kg/m$^3$ and a custom made constant natural curvature $\hat{\kappa}_1$ is fixed at both ends between two concentrically aligned drill chucks separated by a distance  $d = 22$ cm. Our experiment consists of quasi-statically increasing the rotation $\Phi$ at one end and investigating the evolution of equilibrium state with the control parameter $\Phi$. Inset courtesy of Khalid Jawed: Fabrication process of an elastomeric rod. During working time, the PVC tubes containing the silicone-based elastomer are wound around cylinders of external radius $R_e$ and left in that position for $24$ hours. After demolding, it confers to the rod a constant natural curvature $\hat{\kappa}_1(s) \approx 1/R_e$.}
\end{figure}

Our rods are cast by injecting vinylpolysiloxane (VPS), a two-part silicone-based elastomer, into a flexible PVC tube of inner and outer diameters $D_I = 3.1$ mm and $D_O = 5$ mm, respectively. The PVC mold is first wound around a cylinder of external radius $R_e$ and then injected with VPS, which eventually cross-links at room temperature [see inset of Fig.~\ref{fig:experimental_setup}]. After a setting period of $24$ hours, to ensure complete curing of the polymer, the outer flexible PVC pipe is cut to release the inner slender VPS elastic rod with a constant natural curvature $\hat{\kappa}_1(s) = 1/(R_e + D_I/2)$ and a circular cross-section $R = D_I/2 = 1.55$ mm. The rod's second moments of area are $I_1 = I_2 = \pi R^4/4$ and $I_3 = J = \pi R^4/2$. We measure the Young's modulus of the elastomer to be $E = 1000$ KPa, a volumic mass $\rho = 1200$ kg/m$^3$ and a Poisson ratio of $\nu \approx 0.5$, so that its shear modulus is $G = E/2(1+\nu) = 305$ KPa.  

The cast rod ($L = 30$ cm long) is then attached between two horizontal concentric drill chucks of a lathe, separated by a distance $d = 22$ cm. A photograph of the side view of the experiment is presented in
Fig.~\ref{fig:experimental_setup}. The boundary conditions of the rod are set to be rigidly clamped at both ends. For future representation of the rod configurations, we choose the origin of the cartesian frame $\left(x,y,z\right)$ to be located at the clamp at the left extremity of the rod [see Fig.~\ref{fig:experimental_setup}]. The clamp located at the origin, at the curvilinear coordinate $s = 0$, is completely fixed but the other clamp, located at $s = L$, can be rotated with respect to the $y$-axis, thereby imposing a rotation angle $\Phi$ [see Fig.~\ref{fig:experimental_setup}]. Initially, for $\Phi = 0^{\circ}$, we ensure that the sign of the intrinsic curvature $\hat{\kappa}_1$ is such, that the rod naturally bends downwards, in the direction of gravity and that the difference between twist angles, $\kappa_3(s)$, at both ends of the rod, is zero. In that configuration, the equilibrium shape of the clamped rod is close to a planar inflectional elastica as theoretically described in \cite{van2003instability}; the only difference arising from the effects due to gravity which induces a catenary-like configuration. Our writhing experimental protocol then consists of quasi-statically increasing the rotation angle, $\Phi$, at $s=L$ and quantifying the evolution of equilibrium states as a function of this control parameter, $\Phi$. A variety of measurements on the configurations of the rod are performed by imaging the top of the experiment (using a Nikon D90 SLR camera) and subsequent image processing.


\subsection{Modeling of the  boundary conditions for the writhing configuration}

Before we can proceed with a direct comparison between experimental and numerical results, we first need to precise how to account for the specific kinematic boundary conditions and control parameter, $\Phi$, relevant to this specific writhing configuration, in our general numerical framework presented in Section \ref{sec4}. This specific implementation will serve as an example, which, following the series  of procedures and rationale described below, can be extended to other kinematic conditions to solve a variety of  other problems involving thin rods. 
 
Representing the slender elastic rod of Fig.~\ref{fig:experimental_setup} by the discrete 3D Cosserat curve of Fig.~\ref{fig:DIscrete_Cosserat_rod} and applying the numerical method of Section \ref{sec4}, we can write its equilibrium equations in the quadratic form $\vect{f}\left(\vect{u}\right)$ defined in \ref{quadfu}. In the writhing experiment, gravity is the only external force applied to the rod. This gravitational force is represented by the weight of each element, reported at each node. In $\vect{f}\left(\vect{u}\right)$ given in Eq.~(\ref{eq:fufufufu}), we can therefore write $\vect{P}^0 = \vect{P}^L = -1/2\rho g \pi R^2L^2/N^2\vect{e}_z$ at both end nodes and $\vect{p}^i = -\rho g \pi R^2L/N \vect{e}_z$ for all the internal nodes $1 < i < N+1$, where $g  = 9.81$ m/s$^{2}$ is the gravitational acceleration, $N$ is the number of segments and $\vect{e}_z$ is the unit vector in the $z$-direction. Since there are no external moments, we can also write $\vect{m}^j = \vect{0}$ for all the internal elements $1 < j < N$ and $\vect{M}^0 = \vect{M}^L = \vect{0}$ at both ends. 

\begin{figure}[t!]
\includegraphics[width=\textwidth]{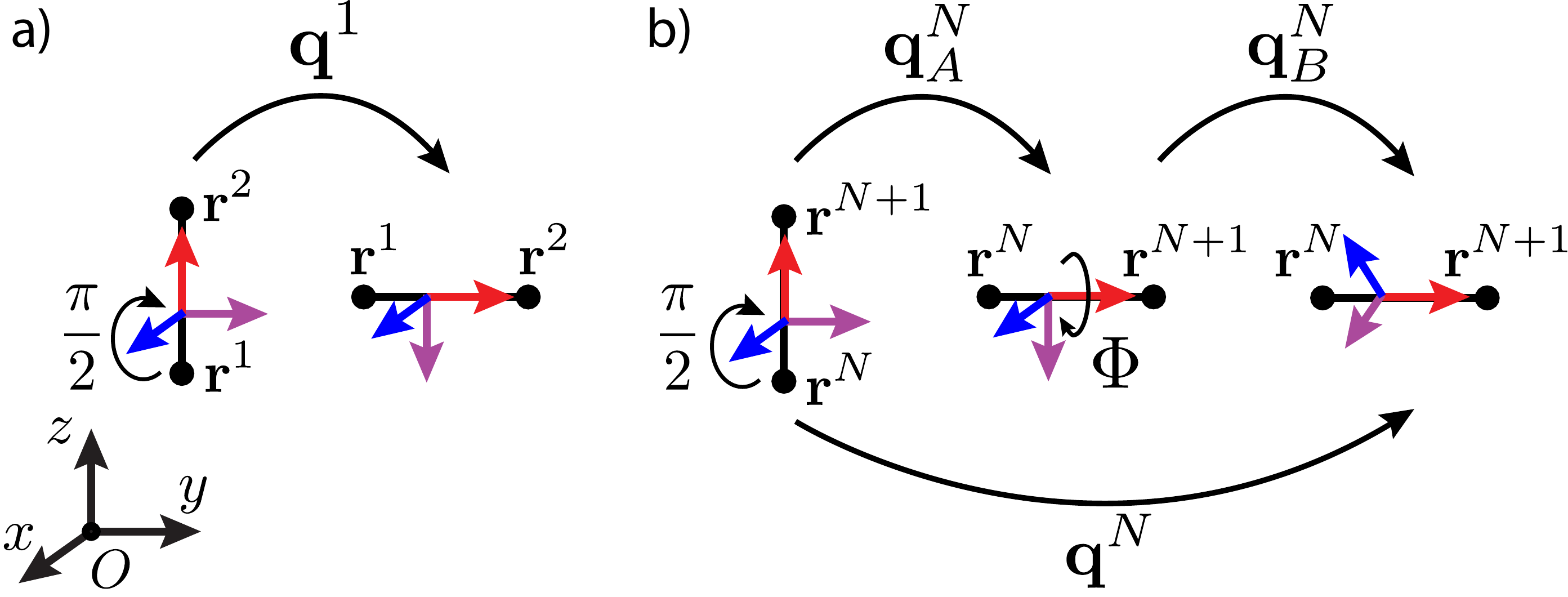}
\caption{\label{fig:Boundary_conditions_writhing} Schematics of the rotational boundary conditions of our writhing experiment. (a) At $s = 0$, we need to orient the first element along the $y$-direction. To do so, we need to impose a rotation of $-\pi/2$ rad around the $x$-axis. (b) At $s = L$, we impose two rotations. We orient the $N^{\text{th}}$ element in the $y$-direction as in (a) and superimpose a rotation $\Phi$, the control parameter, around the $y$-axis.}
\end{figure}

In addition to the mechanical parameters which are general to thin rods, we also need to account for the kinematic boundary conditions and control parameter $\Phi$, specific to the writhing experiment,  which are not included in the general formulation of $\vect{f}\left(\vect{u}\right)$ given in Eq.~(\ref{eq:fufufufu}). At the left extremity of the rod ($s = 0$), the first node must be fixed to the origin and the first element, which is naturally pointing in the $z$-direction so that $\vect{d}_3$ is parallel to $\vect{e}_z$, has to be re-oriented along the $y$-direction [see Fig.~\ref{fig:Boundary_conditions_writhing}.a]. Mathematically, this translates into two functional constraints depending on the positional and rotational degrees of freedom, $\vect{r}^1$ and $\vect{q}_1$ alone,  
\begin{subequations}
\label{eqBCsega0}
\begin{align}
\vect{\mathcal{C}}^0_r\left[\vect{r}^1\right] & = \vect{r}^1 = \vect{0}, \label{eqBCsega0a} \\
\vect{\mathcal{C}}^0_q\left[\vect{q}^1\right] & = \vect{q}^1 - \tp{\left[-\frac{\sqrt{2}}{2} \; 0 \; 0 \; \frac{\sqrt{2}}{2}\right]} = \vect{0}. \label{eqBCsega0b}
\end{align}  
\end{subequations} 
Whereas the first condition~(\ref{eqBCsega0a}) can be physically interpreted as the fixed boundary conditions at $s = 0$, $\vect{r}^1 = \vect{0}$, the second Eq.~(\ref{eqBCsega0b}) is more difficult to interpret due to the lack of direct physical significance of the quaternions. To determine the constraint set in Eq.~(\ref{eqBCsega0b}), we used the relation between Euler's principal geometric quantities and a set of unit quaternions, whose expression was given in Eq.~(\ref{eqeulerparameter}). According to Euler's rotation theorem \cite{kuipers1999quaternions}, the imposed rotation applied to the first segment (shown schematically in Fig.~\ref{fig:Boundary_conditions_writhing}(a) can be represented by a rotation angle of $-\pi/2$ rad around the unit length vector $\vect{b} =  \tp{\left[1 \, 0 \, 0\right]}$. Following the conversion formula of Eq.~(\ref{eqeulerparameter}), the equivalent representation in terms of unit quaternions is given by the rotation $\vect{q}^1 = \tp{\left[-\sin(\pi/4) \; 0 \; 0 \; \cos(\pi/4)\right]}$ provided in Eq.~(\ref{eqBCsega0b}).

At the other end of the rod ($s = L$), where the rotation is being imposed, the last node is fixed at $y = d = 22$ cm. Furthermore, the last element has to be rotated by $-pi/2$ rad around the $x$-axis to re-orient $\vect{d}_3$ along the $y$-direction, as explained above, but we also need to superimpose a rotation $\Phi$ with respect to the $y$-axis, to simulate writhing  [see Fig.~\ref{fig:Boundary_conditions_writhing}.(b)]. Mathematically, these conditions translate into two constraints depending on the positional and rotational degrees of freedom alone, $\vect{r}^{N+1}$ and $\vect{q}_N$, 
\begin{subequations}
\label{eqBCsegaL}
\begin{align}
\vect{\mathcal{C}}^L_r\left[\vect{r}^{N+1}\right] & = \vect{r}^{N+1} - \tp{\left[0 \; d \; 0\right]} = \vect{0} \label{eqBCsegaLa} \\
\vect{\mathcal{C}}^L_q\left[\vect{q}^N\right] & = \vect{q}^N - \frac{\sqrt{2}}{2}\tp{\left[-\cos(\Phi/2) \; \sin(\Phi/2) \; \sin(\Phi/2) \; \cos(\Phi/2)\right]} = \vect{0}. \label{eqBCsegaLb}
\end{align}  
\end{subequations}       
Again, Eq.~(\ref{eqBCsegaLa}) is a direct translation of the positional boundary conditions of our writhing experiment. Eq.~(\ref{eqBCsegaLb}), however, is less intuitive; it is the quaternion representation of the composition of the two rotations shown in Fig.~\ref{fig:Boundary_conditions_writhing}.(b). The first rotation, $A$, has already been treated above and can be described by the quaternion vector $\vect{q}^N_A = \tp{\left[-\sin(\pi/4) \; 0 \; 0 \; \cos(\pi/4)\right]}$. The second re-orientation, $B$, can be represented by a rotation angle of $\Phi$ around the unit length vector $\vect{b} =  \tp{\left[0 \, 1 \, 0\right]}$ and translates as $\vect{q}^N_B = \tp{\left[0 \; \sin(\Phi/2) \; 0 \; \cos(\Phi/2)\right]}$ in the quaternion basis. The total rotation imposed to the $N^{\text{th}}$ element is therefore the composition of the two rotations $A$, then $B$. In terms of quaternions, this rotation $\vect{q}^N$ is represented by the multiplication of the two sets of quaternions $\vect{q}^N_A$ and $\vect{q}^N_B$ and reads $\vect{q}^N = \vect{q}^N_A .\vect{q}^N_B$ which, following the multiplication rule in the quaternion basis \cite{kuipers1999quaternions}, is given in Eq.~(\ref{eqBCsegaLb}).    

To properly account for the boundary conditions in our numerical model introduced in Section \ref{sec4}, in addition to the constraints set by Eqs.~(\ref{eqBCsega0})-(\ref{eqBCsegaL}), we also need to include the quantities $\vect{\mu}^0_r\vect{\nabla}\vect{\mathcal{C}}^0_r$, $\vect{\mu}^L_r\vect{\nabla}\vect{\mathcal{C}}^L_r$, $\vect{\mu}^0_q\vect{\nabla}\vect{\mathcal{C}}^0_q$ and $\vect{\mu}^L_q\vect{\nabla}\vect{\mathcal{C}}^L_q$, into the quadratic vector of equilibrium equations $\vect{f}(\vect{u})$ given in \ref{quadfu}. Doing so involved the introduction of the Lagrange multipliers $\vect{\mu}^0_r$, $\vect{\mu}^L_r$, $\vect{\mu}^0_q$ and $\vect{\mu}^L_q$ in the vector of unknowns $\vect{u}$ of Eq.~(\ref{eqquadraticstatevector}). Physically, $\vect{\mu}^0_r$ and $\vect{\mu}^L_r$ are the forces at each end, written in the Cartesian basis, required to impose the positional boundary conditions Eqs.~(\ref{eqBCsega0a})-(\ref{eqBCsegaLa}) at equilibrium. Similarly, $\vect{\mu}^0_q$ and $\vect{\mu}^L_q$ are a set of quaternions representing the moments at each extremity, necessary to impose the rotational boundary conditions. Finally, these boundary conditions can be accounted for in the stability analysis by including the new Lagrange parameters and their associated constraints in the second-order condition given in Section \ref{secstab} through Eq.~(\ref{eqsecondordercondition}).

Finally, before we are able to solve our modified nonlinear algebraic problem $\vect{f}(\vect{u}) = \vect{0}$, one last important step is needed. The condition given in Eq.~(\ref{eqBCsegaLb}) is not quadratic in terms of the control parameter $\Phi$, and consequently, without further modification, the updated nonlinear vector valued function $\vect{f}(\vect{u})$ would not be adequate to the numerical framework posed in Section \ref{sec4}. Fortunately, there is an appropriate way in the ANM framework to quadratically recast Eq.~(\ref{eqBCsegaLb}). The technique involves adding two new variables,
\begin{subequations}
\label{eq:trigoMANLAB}
\begin{align}
c(a) & = \cos\left(\Phi\left(a\right)/2\right) \label{eq:trigoMANLABa} \\
s(a) & = \sin\left(\Phi\left(a\right)/2\right), \label{eq:trigoMANLABb} 
\end{align}
\end{subequations}
into the vector of unknowns $\vect{u}(a)$. The procedure is based on the introduction of differential equations in terms of the path-parameter $a$ in $\vect{f}(\vect{u}(a))$. Differentiating Eq.~(\ref{eq:trigoMANLAB}) with respect to $a$, the unknowns $(\vect{q}^N,c,s)$ are now solutions of the quadratic algebraic system,
\begin{subequations}
\label{eq:RPhwrithing}
\begin{align}
\vect{0} & = \vect{q}^N(a) - \frac{\sqrt{2}}{2}\tp{\left[-c(a) \; s(a) \; s(a) \; c(a)\right]},\label{eq:RPhwrithinga} \\
0 & = dc(a) + \frac{1}{2}s(a)d\Phi(a), \label{eq:RPhwrithingb} \\
0 & = ds(a) - \frac{1}{2}c(a)d\Phi(a). \label{eq:RPhwrithingc} 
\end{align}
\end{subequations}
To fully integrate Eqs.~(\ref{eq:RPhwrithing}) into the asymptotic numerical framework of Section \ref{ANMmethod}, we now need to perform a minor modification to the identification technique of the power series explained in Eqs.~(\ref{eqnonlinearfunctionalquadratic})-(\ref{eqlinearsystems}). We recall that the fundamental idea behind the ANM is to express the vector of unknowns in a power series of $a$ such that, $\vect{u}(a) = \vect{u}_0 + \sum_{p=1}^{m}a^p\vect{u}_p$, whose differential version reads,
\begin{equation}  
d\vect{u}(a) = \vect{u}_1 + \sum_{p=2}^{m}pa^{p-1}\vect{u}_p.
\label{eqdiffasymptoticdev}
\end{equation}
Substituting Eq.~(\ref{eqdiffasymptoticdev}) into Eq.~(\ref{eq:RPhwrithing}) and identifying the power of $a$ allow us to compute the contributions $\vect{u}_p$ of the semi-analytical vector of unknowns $\vect{u}(a)$, following the same procedure described in in Section \ref{ANMmethod}. We highlight the fact that this method of introducing differential equations in the ANM method is a convenient way to represent complex non-polynomial energy functions in our numerical model, e.g. to represent highly nonlinear phenomena such as contact forces \cite{Karkar2013968}.   

The kinematic boundary conditions, Eqs.~(\ref{eqBCsega0})-(\ref{eqBCsegaL}), and rotational control parameter $\Phi$, Eqs.~(\ref{eq:trigoMANLAB})-(\ref{eqasymptoticdev}), for the writhing problem are all now correctly implemented into our algebraic equilibrium equations $\vect{f}(\vect{u}(a)) = \vect{0}$,  where $\vect{f}(\vect{u})$ is given in Eq.~(\ref{eq:fufufufu}). We can now compute the vector of unknowns, $\vect{u}(a)$, as an asymptotic expansion in terms of the path-parameter, $a$, using the method explained in Section \ref{sec4}, to perform the continuation of the solutions $\vect{u}(a)$ and assess their associated local stability.

\begin{figure}[t!]
\includegraphics[width=\textwidth]{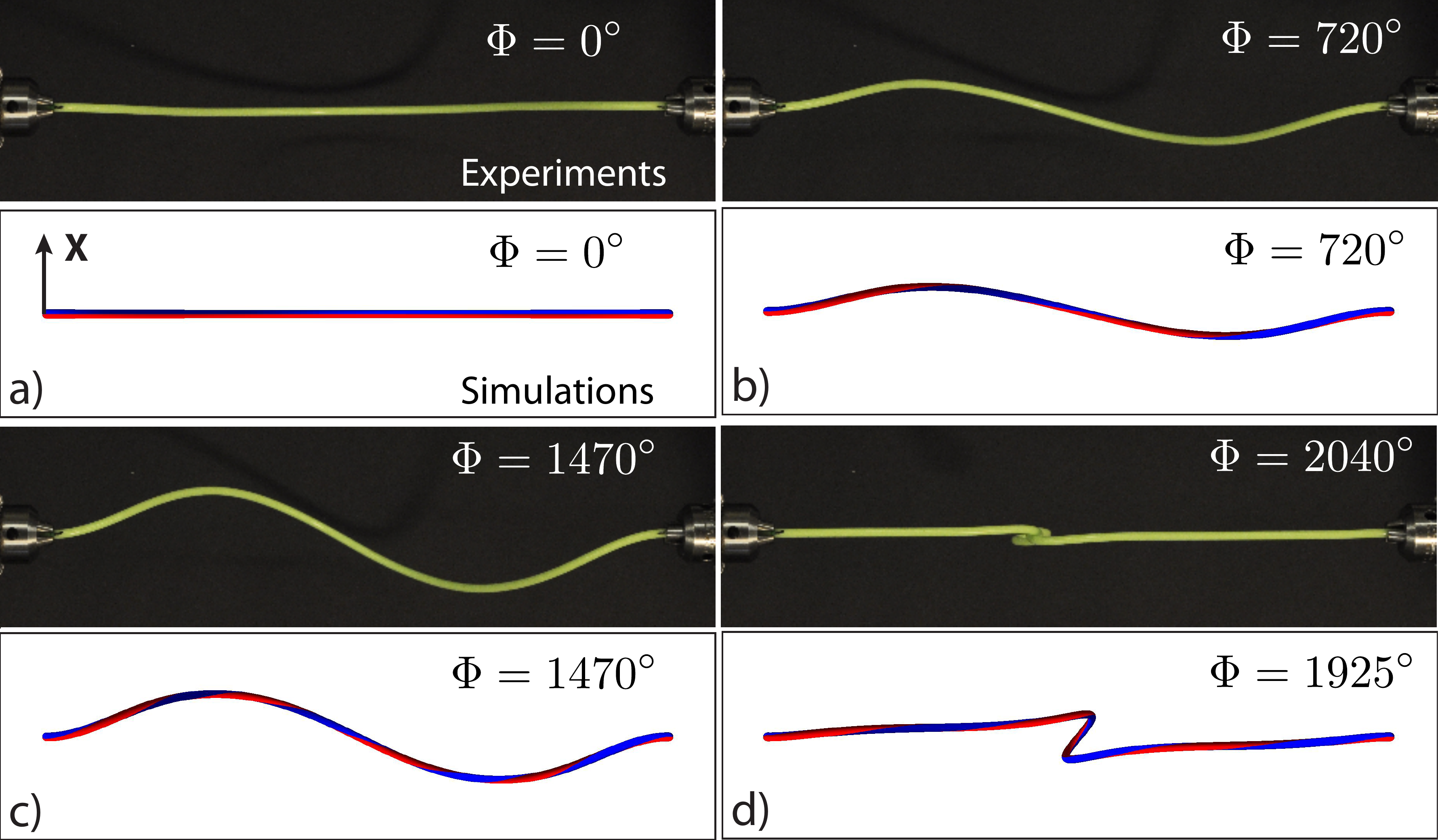}
\caption{\label{fig:Configs_simuexp_k0} Top view of various equilibrium configurations of a straight elastic rod ($\hat{\kappa}_1 = 0$ m$^{-1}$) for increasing values of the rotation angle $\Phi$.  The experimental pictures have a black background and the simulations have a white background.
The simulation results are rendered to visualize twist by using bi-color rods. (a) Planar equilibrium shape at rest for $\Phi = 0^{\circ}$. (b) Out-of-plane configuration for $\Phi = 720^{\circ}$.  (c) Out-of-plane configuration for $\Phi = 1470^{\circ}$. (d) Onset of formation of a plectoneme at the middle of the rod. 
}
\end{figure}

\subsection{Comparison between numerics and experiments}

Having introduced, developed and described our theoretical and computational tools, we proceed by performing a direct comparison between numerics and experiments. In particular, we focus on quantifying the evolution of the equilibrium configurations and associated buckling instabilities, as a function of the control parameter, $\Phi$. We highlight that in this comparison, there are no fitting parameters; all material and geometric parameters of the experiments are independently measured and considered as input variables into the numerics.

\begin{figure}[t!]
\includegraphics[width=\textwidth]{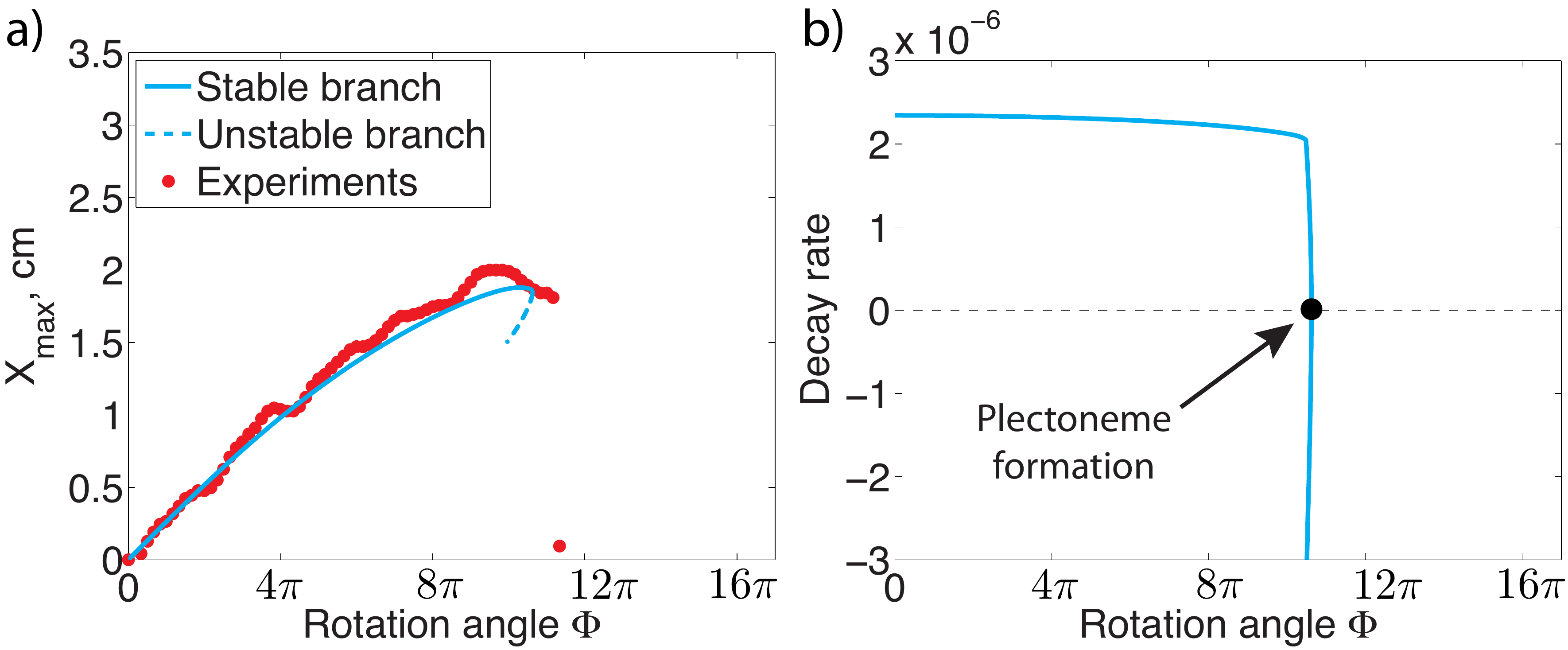}
\caption{\label{fig:Bifurcation_k0} Bifurcation diagram of the straight elastic rod ($\hat{\kappa}_1 = 0$ m$^{-1}$). (a) Evolution of the maximum transverse displacement, $X_{max} = max(abs(X))$ (defined in Fig.~\ref{fig:Configs_simuexp_k0}(a)), with the control parameter $\Phi$. Comparison between experimental results and the semi-analytical branches computed with MANlab. Solid/dashed lines represent stable/unstable branches, respectively. (b) Evolution of the first eigenvalue of the stability problem with control parameter $\Phi$. The critical angle corresponding to the emergence of plectoneme is $\Phi^c_{sim} = 1925^{\circ}$ for the simulations and $\Phi^c_{exp} = 2025 \pm 15^{\circ}$ for the experiments.}
\end{figure}

\begin{figure}[t!]
\includegraphics[width=\textwidth]{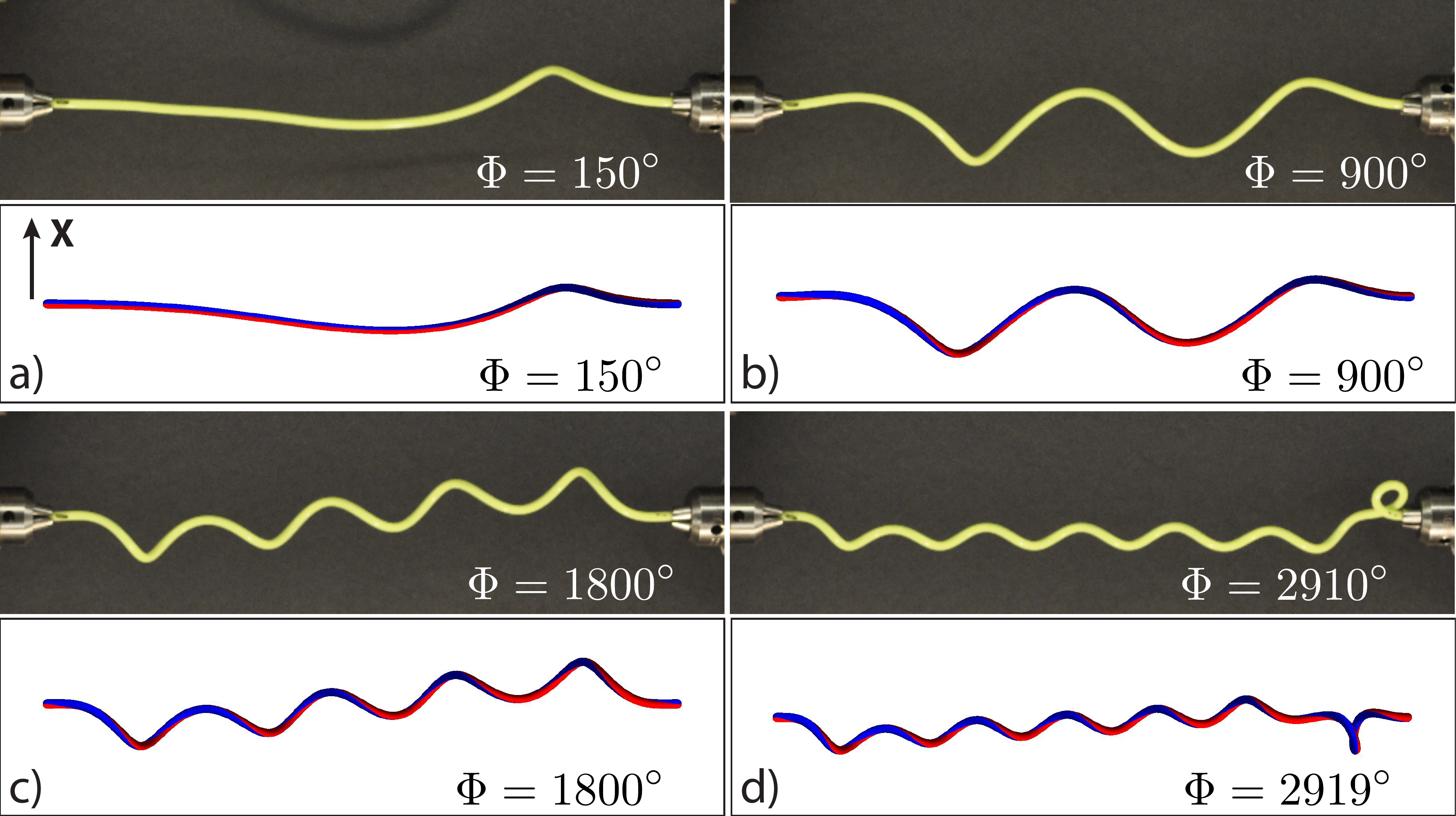}
\caption{\label{fig:Configs_simuexp_k38} Top view of various equilibrium configurations of a curvy elastic rod ($\hat{\kappa}_1(s) = 44.84$ m$^{-1}$) for increasing values of the rotating angle, $\Phi$. 
The experimental pictures have a black background and the simulations have a white background.
The simulation results are rendered to visualize twist by using bi-color rods. (a) Asymmetric out-of-plane equilibrium shape for $\Phi = 150^{\circ}$. (b) One-twist-per-wave configuration for $\Phi = 900^{\circ}$ with three wavelengths between the two clamps.  (c) One-twist-per-wave configuration for $\Phi = 1800^{\circ}$ with five wavelengths between the two clamps. (d) Onset of formation of a plectoneme state at one extremity of the rod. 
}
\end{figure}

\begin{figure}[b!]
\includegraphics[width=\textwidth]{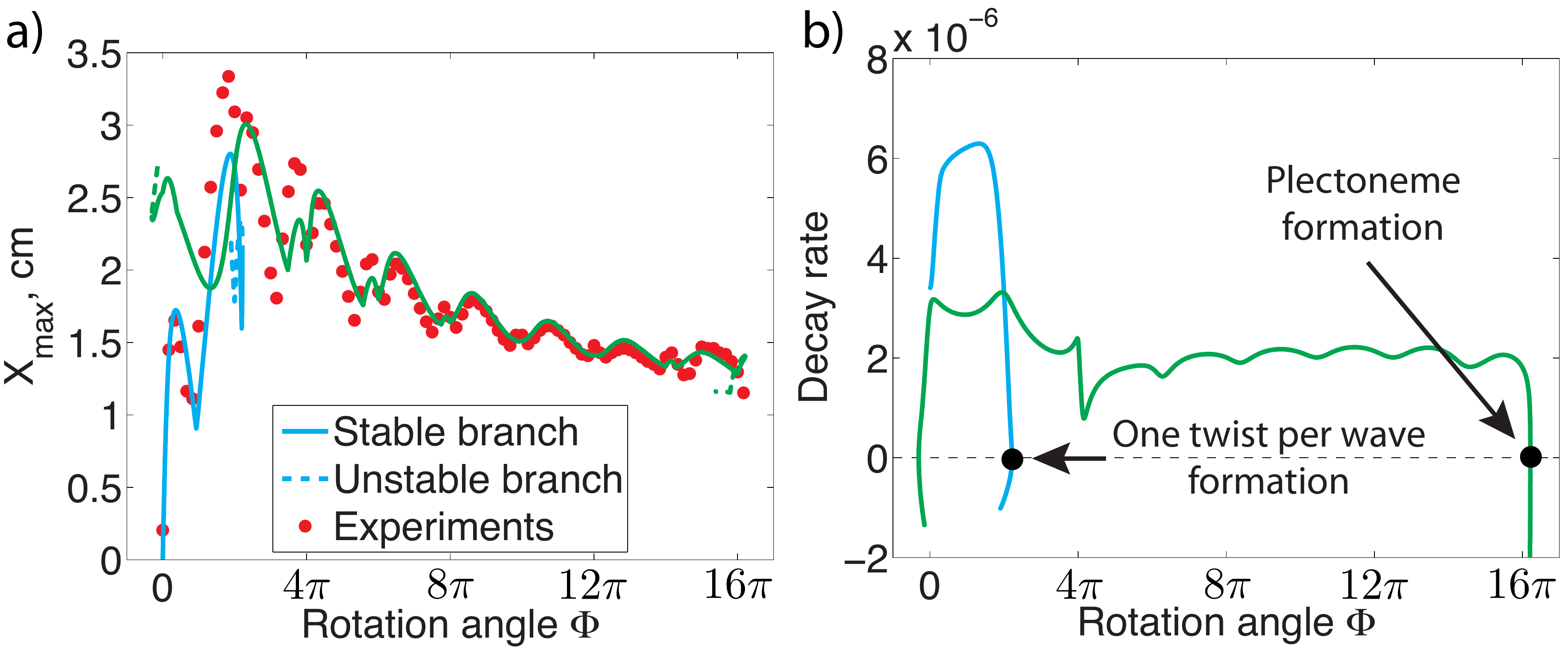}
\caption{\label{fig:Bifurcation_k38} Bifurcation diagram of the curved elastic rod ($\hat{\kappa}_1(s) = 38$ m$^{-1}$). (a) Evolution of the maximum transverse displacement with the control parameter $\Phi$. Comparison between experimental results and the semi-analytical branches computed with MANlab. The presence of a constant natural curvature introduces a new wavy configuration with one-twist-per-wave. Solid/dashed lines represent stable/unstable branches, respectively. (b) Evolution of the first eigenvalue of the stability problem with control parameter $\Phi$. The critical angle corresponding to the emergence of plectoneme is $\Phi^c_{sim} = 2919^{\circ}$ for the simulations and $\Phi^c_{exp} = 2895 \pm 15^{\circ}$ for the experiments.}
\end{figure}

\begin{figure}[b!]
\includegraphics[width=\textwidth]{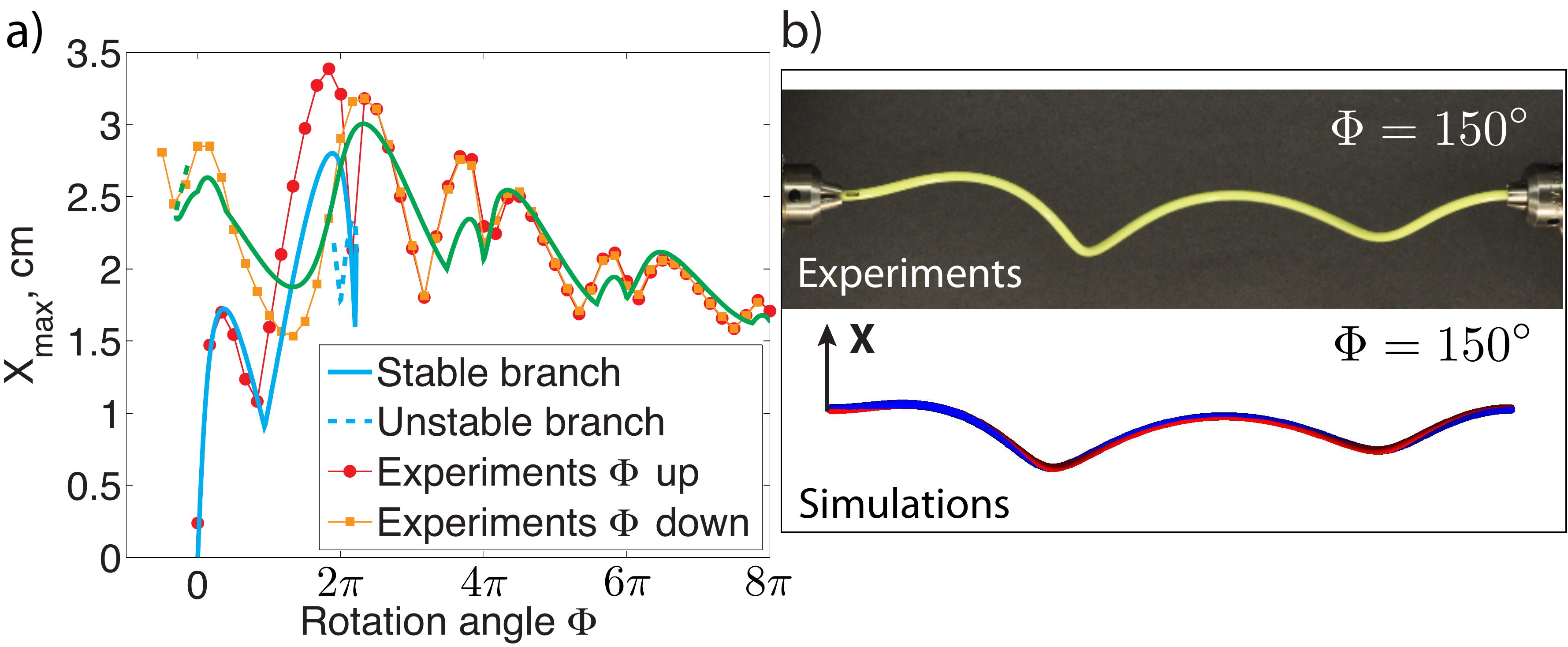}
\caption{\label{fig:Jumps_writhing} Hysteretic behavior of the curved elastic rod ($\hat{\kappa}_1(s) = 44.84$ m$^{-1}$). (a) Evolution of the maximum transverse displacement with the control parameter $\Phi$. Comparison between experimental results obtained for increasing and decreasing $\Phi$ and the semi-analytical branches computed with MANlab. Solid/dashed lines represent stable/unstable branches, respectively. (b) The equilibrium configuration for $\Phi = 150^{\circ}$ (two helices with opposite handedness) is different from the configuration in Fig.~\ref{fig:Configs_simuexp_k38}(a) for the same control parameter $\Phi$, which is significative of multi-stability.}
\end{figure}

In Figs.~\ref{fig:Configs_simuexp_k0}, ~\ref{fig:Configs_simuexp_k38} and \ref{fig:Jumps_writhing}b), we compare the top view of some representative experimental and numerical equilibrium shapes for a straight rod ($\hat{\kappa}_1(s) = 0$ m$^{-1}$) and a naturally curved rod ($\hat{\kappa}_1(s) = 44.84$ m$^{-1}$). From these images, we measure the maximum transverse displacement of the rod, $X_{\mathrm{max}}$, 
in the $(x,y)$ plane (top view) as a function of the rotation angle, $\Phi$, which is treated as a control parameter. Experimentally, the quantity $X_{max}$ was measured from image analysis of the digital images taken by the camera located above the apparatus. Using these quantities, we then construct the bifurcation diagrams presented in Figs.~\ref{fig:Bifurcation_k0}(a), \ref{fig:Bifurcation_k38}(a) and  \ref{fig:Jumps_writhing}(a) (for experiments and numerics), for the straight and curved rods, respectively. We also analyze the stability of the equilibrium state by calculating the first eigenvalue of the stability problem as a function of the control parameter, $\Phi$, and the results are plotted in Figs.~\ref{fig:Bifurcation_k0}(a) and \ref{fig:Bifurcation_k38}(a) (for numerics). In order to quantitatively validate our ANM continuation technique, the  semi-analytical numerical curves (lines) are superposed onto the experimental results (data points, every $\Phi = 30^{\circ}$), for the same value of the control parameter. We highlight, once again, that there are no fitting parameters involved in this comparison; all quantities are measured in the experiments, independently from the numerics. The excellent quantitative agreement between experiments and numerics illustrates the sticking predictive power of our framework.

We now comment on the experimental and numerical results in more detail, focusing first on the case of the straight rod ($\hat{\kappa}_1 = 0$ m$^{-1}$), the results of which we plotted in Figs.~\ref{fig:Configs_simuexp_k0} and~\ref{fig:Bifurcation_k0}. Initially, for $\Phi = 0^{\circ}$, the rod exhibits a planar equilibrium shape lying in the $(y,z)$ plane due to the effect of gravity. This equilibrium configuration is calculated using a classic Newton-Raphson algorithm \cite{riks1979incremental} and taken as the initial fixed point $\vect{u}_0$, the solution of our equilibrium equations $\vect{f}(\vect{u}(a)) = \vect{0}$ [see Fig.~\ref{fig:Configs_simuexp_k0}.(a)]. When the rotation angle $\Phi$ is increased, this initial planar shape evolves smoothly into an out-of-plane configuration, symmetric to the $(y,z)$ plane, with an amplitude that grows due to an increasing internal twist [see Fig.~\ref{fig:Configs_simuexp_k0}.(b)-(c)]. At a critical value of the rotation angle, $\Phi^c$, the out-of-plane shape loses stability and the rod buckles into a \emph{plectoneme} state \cite{thompson1996helix,van2003instability}: a highly localized structure corresponding to a two-start right-handed helix with terminal loops. Beyond this point, our numerical model is no longer able to reproduce the rod's configurations since they involve self-contact which is not included in our description. To further quantify this process, in Fig.~\ref{fig:Bifurcation_k0} we plot the maximum transverse displacement of the rod, $X_{max}$, as a function of the imposed rotation angle, $\Phi$. Across the full range of $\Phi$ explored, the experimental data plotted in Fig.~\ref{fig:Bifurcation_k0} is in excellent quantitative agreement with the numerical prediction.

It is remarkable that the instability threshold for the formation of the plectoneme, $\Phi^c$, is also well recovered by our local stability analysis showed in Fig.~\ref{fig:Bifurcation_k0}.(b), where we plot the evolution of the first eigenvalue, $\sigma_1$, as a function of $\Phi$. With no fitting parameters, the predicted critical threshold $\Phi^c_{sim} = 1925^{\circ}$ is in excellent agreement (within $5\%$) with the experimental results $\Phi^c_{exp} = 2025 \pm 15^{\circ}$. To plot the semi-analytical bifurcation diagram of Fig.~\ref{fig:Bifurcation_k0}, we computed $80$ solution vectors, $\vect{u}(a)$, expressed in terms of power series expansions at the order $m = 20$, as given in Eq.~(\ref{eqasymptoticdev}). Using a desktop computer with a standard processor (at the time of writing) of $2.71$ Ghz and $2.75$ Gb of RAM, the computation required $9$ seconds to determine one asymptotic series. This represents a total running time of approximately $12$ minutes to simulate the full problem, using MANlab. Note that this computational time is mostly due to the $80$ inversions of the Jacobian matrix needed to solve the linear systems given in Eq.~(\ref{eqlinearsystems}), which are of size $1415 \times 1415$ for $N = 100$ elements. These computations could be made even more efficient by using a dedicated solver such as the ones offered by traditional finite element codes but the time optimization of the ANM algorithm, which has been investigated \cite{cochelin1994asymptotic}, is beyond the scope of this paper.

Interestingly, for the case of the naturally curved rod ($\hat{\kappa}_1(s) = 44.84$ m$^{-1}$) the evolution of equilibrium configurations with the rotation angle $\Phi$ is qualitatively different from the case of the naturally straight elastic rod, as shown in Figs.~\ref{fig:Configs_simuexp_k38}, ~\ref{fig:Bifurcation_k38} and \ref{fig:Jumps_writhing}. Once again, the qualitative and quantitative agreement between the experimental and numerical results is remarkable. By introducing the natural curvature $\hat{\kappa}_1(s)$, the previously symmetric out-of-plane solutions obtained for the case of straight rods, become asymmetric with respect to the $(y,z)$ plane. For small rotation angles $\Phi$, the initial planar shape exhibits an asymmetric out-of-plane configuration due to the competition between the imposed internal twist and the intrinsic twist naturally imposed by $\hat{\kappa}_1(s)$ [see Fig.~\ref{fig:Configs_simuexp_k38}.(a)]. Above $\Phi \approx 400^{\circ}$, our results confirm that the rod, jumps into a one-twist-per-wave mode due to the presence of natural curvature, as originally reported in \cite{champneys1997spatially}. In this configuration, the number of waves is equal to the number of twists stored in the rod as shown in Figs.~\ref{fig:Configs_simuexp_k38}.(b)-(c). For a critical rotation angle $\Phi^{c}$, a plectoneme forms, superimposed onto the one-twist-per-wave equilibrium state which is no longer stable. It is interesting to note that, for the naturally curved rod, the plectoneme is located at one extremity of the rod rather than at its center [see Fig~\ref{fig:Configs_simuexp_k38}.(d)], as found above for the straight rod.

For the naturally curved rod, our continuation method is able to robustly and efficiently follow the equilibrium branches across the full range of considered rotation angles, $\Phi$, exhibiting excellent agreement with the experimental data, as shown in Fig.~\ref{fig:Bifurcation_k38}. For this second test case with $\hat{\kappa}_1(s) = 44.84$ m$^{-1}$, we  used $N = 200$ elements and we computed $115$ asymptotic expansions $\vect{u}(a)$ to determine the full semi-analytical bifurcation diagram, leading to a computational time of approximately $35$ minutes. Again, the instability threshold for the onset of a plectoneme is well recovered by our local stability analysis showed in Fig.~\ref{fig:Bifurcation_k38}(b). In this case of a more complicated bifurcation diagram, it is remarkable that the predicted $\Phi^c_{sim} = 2919^{\circ}$ is within $1\%$ of the experimentally measured value of $\Phi^c_{exp} = 2895 \pm 15^{\circ}$. Counterintuitively, we find that imparting a constant natural curvature to our rods 
(essentially adding a finite imperfection to the stress-free configuration) results in postponing, by approximately $43\%$, the emergence of the plectoneme instability (often synonymous with failure in practical systems). To our knowledge, this interesting novel phenomenon has thus far been overlooked in the literature and deserves further investigation. A systematic study to quantify and rationalize the influence of natural curvature on the writhing of a slender elastic rod is beyond the scope of this paper, but is an aspect which we plan investigate in future work. 

Another interesting feature in the writhing of the naturally curved rod is the hysteric behavior observed for small values of the rotation angle $\Phi$, before entering the one-twist-per-wave regime, as illustrated in Fig.~\ref{fig:Jumps_writhing}. Initially, when we  increased $\Phi$, the material twist originally stored due to the natural curvature $\hat{\kappa}_1(s)$ is released until the rod jumps to the one-twist-per-wave mode at $\Phi \approx 400^{\circ}$. If we then decrease $\Phi$, the observed equilibrium configurations are not the asymmetric out-of-plane shapes we previously encountered as illustrated in Fig.~\ref{fig:Configs_simuexp_k38}(a). Instead, Fig.~\ref{fig:Jumps_writhing}(b) shows an inversion of helix handedness, known as perversion \cite{goriely1998spontaneous} and described as two helices with opposite handedness. If we decrease $\Phi$ even further to negative values, the rod jumps back on the previous stable equilibrium branches where the rod twists with the same handedness. In the regime of small rotations  ($\Phi < 400^{\circ}$), our system is metastable; for the same control parameter $\Phi$, the rod can exhibit two different configurations depending on the loading path. Once again, our continuation method correctly predicts the different equilibrium states and stability threshold of this complex hysteric behavior as shown in Fig.~\ref{fig:Jumps_writhing}.  Moreover, this highly nonlinear feature emphasize the ability of our numerical technique to follow the equilibrium branches and stability of slender elastic rods independently of the complexity of the bifurcation diagram.

\section{Conclusions and perspectives}

We have presented an original theoretical and computational framework to follow the equilibria and stability of slender elastic rods. In our model, we account for the elastic energy due to changes of material curvatures and twist, as well as the work of external forces and moments, under the assumption that the rod is inextensible and unshearable. The main novel feature in our continuation method is the use of quaternions to represent rotations. This formulation allows for the 3D kinematics to be treated in a geometrically-exact way and result in equilibrium equations that are, at most, quadratic with respect to the state variables. We have shown that this quadratic recast of geometric nonlinearities is particularly well suited for implementation into an asymptotic numerical method. This powerful perturbation technique provides access to branches of equilibrium solutions in the form of successive portions of power series expansion by consecutively solving a set of linear systems. The equilibrium branches can thereby be followed and their stability evaluated as a function of the control parameters. Finally, we have challenged and validated our computational framework by considering the specific problem of writhing of a thin rod and contrasting our numerical implementation with our own experimental results, finding excellent quantitative agreement between the two. We were able to successfully and accurately calculate the geometrically-nonlinear configurations of the rods, as well as the critical thresholds for instability, with remarkable predictive power. We note that our continuation algorithm is able to address regions of multi-stability and hysteresis, as in the regime of low rotation angles, when increasing or decreasing the control parameter.

A potential extension of this work would be to incorporate other additional mechanical ingredients into our model such as internal stretching, hydrostatic loading and contact forces arising, for instance, due to self-contact or when the rod interacts with external boundaries.  One technical requirement in order to be able to introduce new energy terms in our description is that the resulting equilibrium equations have to be quadratic to match the present ANM framework. However, we have shown that even some some cases of non-polynomial functions can easily be reduced to a quadratic form by introducing a limited number of new variables in the vector of unknowns in a process we call recasting.  Otherwise, the introduction of these new features can be readily accomplished, as long as they derive from a potential energy since continuation methods only apply for conservative systems, where an equilibrium can be found.

We have developed a predictive computational framework to tackle the simulations of extreme displacements and rotations in slender elastic rods. Our novel method is relatively simple to implement, robust, accurate, flexible and computationally efficient. We hope that this technique will be invaluable in problems that demand the predictive understanding of the stability, buckling, snap-through and other complex mechanical phenomena intrinsic to the extreme deformation of slender elastic rods, whose timely revival is highly relevant in a variety of currently open problems in both nature and technology.

\section*{Acknowledgements}

We thank Basile Audoly for introducing us the breathtaking world of quaternions and Matt Metlitz for help with the writhing experiments. We are grateful to the support by the National Science Foundation (CMMI-1129894) and Schlumberger-Doll Research. Arnaud Lazarus acknowledges funding from a Battelle-MIT postdoctoral fellowship.  

\clearpage
\newpage

\appendix

\section{Quadratic form of the vector of equilibrium equations}\label{quadfu}

The quadratic form of the $14N$-dimensional nonlinear vector valued function $\vect{f}\left(\vect{u}\right)$ given in Eq.~(\ref{eqnonlinearfunctionalquadratic}) and representing the equilibrium equations~(\ref{eqequilibriumforcediscrete})-(\ref{eqconstraintsdiscrete}) of our inextensible and unshearable slender elastic rod, can be written as

\begin{equation}  
\vect{f}(\colF{\vect{u}}) =
\left\{\begin{array}[c]{cccccr}
\vspace{0.3cm}

\vect{P}^0 & + & \colF{\vect{\mu}^i} & + & \vect{0} & \text{for $i = 1$} \\
\vspace{0.3cm}

\vect{p}^iL/N & + & \colF{\vect{\mu}_i} - \colF{\vect{\mu}^{i-1}} & + & \vect{0} & \text{for $i = [1\, N]$} \\
\vspace{0.3cm}

 \vect{P}^L & - & \colF{\vect{\mu}^{i-1}} & + & \vect{0} & \text{for $i = N+1$} \\
\vspace{0.3cm}

\vect{M}^0 & + & \vect{0} & + & \begin{array}{c} 2\sum\nolimits_{k=1}^3 \mat{B}_k \colF{\vect{q}^{j+1}}\colF{G^j_k}  \\ + \left(\colF{\mat{D}(\vect{q}^j)\vect{\mu}^j} - \colF{\alpha^j\vect{q}^j}\right)   \end{array} & \text{for $j= 1$} \\
\vspace{0.3cm}

\vect{m}^j & + & \vect{0} & + & \begin{array}{c} 2\sum\nolimits_{k=1}^3 \mat{B}_k\colF{\vect{q}^{j-1}G^{j-1}_k} \\ - 2\sum\nolimits_{k=1}^3 \mat{B}_k \colF{\vect{q}^{j+1}G^{j}_k}   \\ + 2\left(\colF{\mat{D}(\vect{q}^j)\vect{\mu}^j} - \colF{\alpha^j\vect{q}^j}\right) \end{array} & \text{for $j= [1 \, N-1]$} \\
\vspace{0.3cm}

-\vect{M}^L & + & \vect{0} & + & \begin{array}{c} 2\sum\nolimits_{k=1}^3 \mat{B}_k \colF{\vect{q}^{j-1}G^{j-1}_k} \\ + 2\left(\colF{\mat{D}(\vect{q}^j)\vect{\mu}^j} -\colF{\alpha^j\vect{q}^j}\right) \end{array} & \text{for $j= N$} \\
\vspace{0.3cm}

1 & + & \vect{0} & - & \colF{\vect{q}^j}\colF{\vect{q}^j} & \text{for $j= [1 \, N]$} \\
\vspace{0.3cm}

\vect{0} & + & \colF{\vect{r}^{j+1}} - \colF{\vect{r}^j} & - & L/N\colF{\vect{d}_3^j(\vect{q}^2}) & \text{for $j= [1\, N]$} \\
\vspace{0.3cm}

E_1I_1\hat{\kappa}_1^j & + & \colF{G_1^j}L/N & - & \begin{array}{c} E_1I_1\mat{B}_1\left(\colF{\vect{q}^j}+\colF{\vect{q}^{j+1}}\right) \\ .\left(\colF{\vect{q}^{j+1}}-\colF{\vect{q}^j}\right) \end{array} & \text{for $j= [1\, N-1]$} \\
\vspace{0.3cm}

E_2I_2\hat{\kappa}_2^j & + & \colF{G_2^j}L/N & - & \begin{array}{c} E_2I_2\mat{B}_2\left(\colF{\vect{q}^j}+\colF{\vect{q}^{j+1}}\right) \\ .\left(\colF{\vect{q}^{j+1}}-\colF{vect{q}^j}\right) \end{array} & \text{for $j= [1\, N-1]$} \\
\vspace{0.3cm}

E_3I_3\hat{\kappa}_3^j & + & \colF{G_3^j}L/N & - & \begin{array}{c} E_3I_3\mat{B}_3\left(\colF{\vect{q}^j}+\colF{\vect{q}^{j+1}}\right) \\ .\left(\colF{\vect{q}^{j+1}}-\colF{\vect{q}^j}\right) \end{array} & \text{for $j= [1\, N-1]$} 
\end{array} 
\right.
\label{eq:fufufufu}
\end{equation}
where the expression of the director $\vect{d}_3^j(\vect{q}^2)$, the projection matrix $\mat{D}(\vect{q}^j)$ and the unknown variables $\vect{u}$ can be found in  Eq.~(\ref{eqrotationmatrixdiscrete}), Eq.~(\ref{eqderived3discrete}) and Eq.~(\ref{eqquadraticstatevector}), respectively. Note that these algebraic equations are the general form of the equilibrium equations of the inextensible elastic rod under external forces and moments and do not account for particular kinematic boundary conditions or control parameter which can vary depending on the problem under study. An example of he additional equations needed to impose some fixed boundary conditions and a rotation angle as control parameter for the case of writhing of a rod is given in Section \ref{sec5}.

\newpage
\bibliography{Bibliobus}

\end{document}